\documentclass[twocolumn,         % Format : preprint, twocolumn
               showkeys,showpacs,            % Pacs : showpacs, noshowpacs
               preprintnumbers,longbibliography,     % Preprint: preprintnumbers,
               aps,                 % Society: ...
               prd,          	    % Journal Style : pra, prb, prc, prd, pre,
               letterpaper,             % Size : a4paper, ...
               nofootinbib,         % Footnote: footinbib, nofootinbib
               tightenlines,        % Remove additional spaces in a line
               floats,floatfix      % Floating pictures and tables
               ]{revtex4-1}
\usepackage{dcolumn}   % needed for some tables
\usepackage{bm}        % for math
\usepackage{float}
\usepackage{physics}
\usepackage{graphicx,color}
\usepackage{subfigure}
\usepackage{latexsym}
\usepackage{amsmath,amssymb}        
\usepackage[colorlinks=true,linkcolor=blue,citecolor=blue]{hyperref}
\usepackage{mathrsfs}
\usepackage{comment}
\usepackage{soul}
\usepackage{cancel}
\definecolor{purple}{rgb}{0.58,0.0,0.83}
\definecolor{orange}{rgb}{1,0.5,0}
\DeclareSymbolFontAlphabet{\mathrsfs}{rsfs}
\DeclareMathAlphabet{\mathcal}{OMS}{cmsy}{m}{n}

\begin{document}

% -----> TITLE 

\title{Spherical solutions of the Schr\"odinger-Poisson system with core-tail structure}

% ----->   AUTHORS   <-----

\author{Iv\'an \'Alvarez-Rios}
\email{ivan.alvarez@umich.mx}
\affiliation{Instituto de F\'{\i}sica y Matem\'{a}ticas, Universidad
              Michoacana de San Nicol\'as de Hidalgo. Edificio C-3, Cd.
              Universitaria, 58040 Morelia, Michoac\'{a}n,
              M\'{e}xico.}

\author{Francisco S. Guzm\'an}
\email{francisco.s.guzman@umich.mx}
\affiliation{Instituto de F\'{\i}sica y Matem\'{a}ticas, Universidad
              Michoacana de San Nicol\'as de Hidalgo. Edificio C-3, Cd.
              Universitaria, 58040 Morelia, Michoac\'{a}n,
              M\'{e}xico.}

% --->   DATE

\date{\today}

% -----> ABSTRACT

\begin{abstract}
We construct spherically symmetric equilibrium solutions of the Schr\"odinger-Poisson (SP) system of equations with a core-tail structure that could serve as models of Fuzzy Dark Matter (FDM) halos. The core is assumed to be a solitonic ground state equilibrium configuration of the SP equations, and the tail is integrated from a transition radius onwards. The total mass of the system parametrizes the family of solutions and constrains the tail density profile. The tail has a radial velocity profile, whereas the core is stationary. We investigate the evolution of these equilibrium configurations and find that the tail initially perturbs the core, and consequently, the whole solution oscillates around a virialized solution that we call 'relaxed', whose average also has a core-tail structure. We measure the departure of the relaxed configuration from the equilibrium solution in order to estimate the utility of the latter. We also find that the core-halo scaling relation of equilibrium configurations has an exponent $\alpha=1/3$, whereas relaxed configurations exhibit a scaling with $\alpha=0.54$.
\end{abstract}

% ----->   PACS

%\pacs{Numbers here}
%07.05.Tp Computer modeling and simulation
%07.05.Mh Neural networks, fuzzy logic, artificial intelligence
%05.45.Tp Time series analysis
%04.30.-w Gravitational waves

% ----->   MAKETITLE   <-----

\maketitle

% ---------------------------------------------
% ----->     INTRODUCTION.    <-----
% ---------------------------------------------
\section{Introduction}

Ultralight Bosonic Dark Matter or Fuzzy Dark Matter (FDM) is a candidate currently under study because it shows interesting properties. Its mass is of order $10^{-23}-10^{-20}$eV, which implies a minimum size of the structures it forms due to its large de Broglie wavelength (see e. g. the reviews \cite{Suarez:2013,Hui:2021tkt,ElisaFerreira,Chavanis2015}). The state of the art on the subject involves sophisticated structure formation simulations (e.g. \cite{Schive:2014dra,Mocz:2017wlg,May_2021}), based on the solution of the equations that rule the dynamics of this type of matter, the Schr\"odinger-Poisson (SP) system of equations, that have captured two essential features of the model, namely, that it behaves like Cold Dark Matter (CDM) at large scales, whereas at local galactic scales it  forms clumps with a  core-tail structure, where the core is the ground state solution of the SP system \cite{Ruffini:1969,GuzmanUrena2004} and the tail has a Navarro-Frenk-White (NFW) decay \cite{Chan_2022}. %, as found in CDM structure formation simulations \cite{NFW}.

The characterization of core-tail profiles obtained in Structure Formation Simulations (SFS) is a subject of current interest and a garden variety of mass-energy combinations and scaling relations arise \cite{Chan_2022}. Modeling core-halo structures resulting from SFS includes self-consistent constructions of the wave function, that suits the density profiles in terms of multimode expansion of the wave function associated to virialized structures, as first seen in \cite{PhysRevD.97.103523}. Nevertheless SFS are costly, which makes the study of galactic phenomenology difficult or unaffordable. For instance, the construction of a catalog of FDM halos is limited to the type of profiles obtained in SFS, which unavoidably cannot have all possible combinations of core-halo mass ratios, nor all the parameters of the core and NFW density functions.

In order to overcome this limitation and at the same time avoid expensive SFS, there are some strategies to construct core-tail FDM configurations without the need of performing SMS. One of them consists in the evolution of multi-core mergers that end up with the desired core-tail type distribution (e. g. \cite{Luna2023}); in the end, the merger of a small number of halos requires less computer resources than SFS and therefore the construction of FDM halos is affordable. Another method consists in the {\it ab initio} construction of the wave function describing a desired density profile, as in \cite{YavetzLiHui2022}, where solutions are constructed with a solitonic core and the tail consisting of the superposition of modes that replicate core and tail profiles at will. In the analysis of \cite{PhysRevD.97.103523}, a similar multi-mode construction was needed and constructed with the aim of building virialized structures that could be compared with density profiles obtained in SFS simulations. These are strategies that allow the construction of realistic galactic halos, in turn help to study galactic dynamics avoiding the cost of halos obtained from SFS. 

In this paper we propose a different method to construct {\it equilibrium} core-tail configurations, which are solutions of the SP system of equations assuming spherical symmetry and quantum-hydrostatic equilibrium. These configurations cover the domain $[0,r_{max}]=[0,r_t]\cup[r_t,r_{max}]:={\rm core~region}\cup {\rm tail~region}$, where $r_t$ is a transition radius between the {\it core} and {\it tail} regions. In the {\it core} we assume there is a solitonic solution, and from $r_t$ on we integrate the stationary SP equations assuming some boundary conditions at $r_t$ and $r_{max}$. The family of solutions is parametrized by the total mass of the configuration and the tail density-profile adjusts itself to fulfill the total mass value.

Non-trivial {\it equilibrium} solutions are found for non-zero radial velocity of the FDM in the tail region, which compromises their stability. In order to study the effects of the tail velocity we evolve the {\it equilibrium} solutions and find that after an initial transient the density redistributes and forms a new configuration that we call {\it relaxed} solution, because it results from the evolution of an equilibrium solution that is somehow in a tense equilibrium state. The {\it relaxed} solution oscillates near a virialized state and in average it also has the core-tail structure. This is not a new effect, it also happens in many-multimode solutions constructed in \cite{YavetzLiHui2022}, where the assumption of stationarity is only an approximation allowing the construction of initial states, although after evolution the configuration evolves and relaxes around a new redistributed configuration that oscillates in time.
 
A further analysis includes the rotation curve (RC) associated to both {\it equilibrium} and {\it relaxed} configurations. We measure how the RC of the latter departs from that of the former. We find that the departure is of a few percent even for core-halo mass ratios of order $\sim 0.2$, which corresponds to tail dominating configurations, and while more core-dominating configurations are the departure should be even smaller.

Finally, another property we explore is the core-halo scaling relation, and find that {\it equilibrium} solutions obey the scaling with an exponent $\alpha\sim 1/3$, whereas {\it relaxed} solutions are better fitted with $\alpha\sim 0.54$, consistently with previous analyses and formation processes. 

The paper is organized as follows. In Section \ref{sec:solutions} we describe the construction of the solutions, in Section \ref{sec:evolution} we show how these evolve and their average behavior. Finally in Section \ref{sec:conclusions} we draw some conclusions.

%%%%%%%%%%%%%%%%%%%%%%%
\section{Solutions}
\label{sec:solutions}

The SP equations in dimensionless units describing the dynamics of the bosonic dark matter is written as

\begin{eqnarray}
i\partial_t \Psi = -\dfrac{1}{2}\nabla^2\Psi + V\Psi, \label{eq:Psi} \\
\nabla^2V = |\Psi|^2 - \expval{|\Psi|^2}, \label{eq:V} 
\end{eqnarray}

\noindent where

\begin{equation}
\expval{|\Psi|^2} = \dfrac{1}{V_D}\int_D |\Psi|^2 d^3x,
\end{equation}

\noindent is the average mass density, calculated in the numerical spatial domain $D$ with volume $V_D$. We translate this system into the quantum-hydrodynamic version in the Madelung frame by using the wave function transformation  $\Psi = \sqrt{\rho} e^{iS}$ \cite{Bohm1952a}, so that the system (\ref{eq:Psi})-(\ref{eq:V}) reduces to equations for mass density $\rho$ and phase of the wave function $S$:

\begin{eqnarray}
\partial_t \rho + \div(\rho \vec{v}) = 0, \label{eq:density} \\
\partial_t S + \frac{1}{2} |\vec{v}|^2 + Q + V = 0, \label{eq:phase} \\
\nabla^2 V = \rho-\expval{\rho}, \label{eq:potential}
\end{eqnarray}

\noindent where $\vec{v} := \grad S$ is interpreted as the velocity vector field of the quantum fluid subject to the influence of the quantum potential $Q = -\frac{1}{2} \frac{\nabla^2 \sqrt{\rho}}{\sqrt{\rho}}$ and the gravitational potential $V$ sourced by the density $\rho$ itself.\\

Assuming spherical symmetry and time-independence of density, we use the {\it ansatz} for the phase $S(t,r) = -V_0 t + \bar{S}(r)$, where $r $ is the radial coordinate, and the system reduces to the following set of ordinary differential equations:

\begin{eqnarray}
\frac{1}{r^2} \frac{d}{dr}(r^2 \rho v) = 0, \label{eq:equilibrium rho}\\
\frac{1}{2} v^2 + Q + V = V_0, \label{eq:equilibrium vr} \\
\frac{1}{r^2} \frac{d}{dr} \left(r^2 \frac{dV}{dr}\right) = \rho-\expval{\rho}, \label{eq:equilibrium V}
\end{eqnarray}

\noindent where the radial velocity is $v :=  \frac{d\bar{S}}{dr}$. The solitonic core of structures found in simulations \cite{Schive:2014dra,Mocz:2017wlg}, in average coincides with ground state equilibrium solutions constructed assuming $v=0$ and asymptotic isolation $\expval{\rho}=0$ as described in \cite{AlvarezGuzmanMadelung}, which are equivalent to those in the Schr\"odinger frame \cite{Ruffini:1969,GuzmanUrena2004}. These solutions are approximated with the solitonic density profile obtained empirically from structure formation simulations \cite{Schive:2014dra,Mocz:2017wlg}:

\begin{equation}
\rho_{core}(r) = \rho_{0,core} \left[1+0.091 \left(\frac{r}{r_c}\right)^2\right]^{-8},\label{eq:core density}
\end{equation}

\noindent where $V_0 \approx 0.6922$, $\rho_{0,core}\approx 2.9208r_c^{-4}$ with $r_c \approx 1.3073$.

Similarly to the construction in \cite{PhysRevD.97.103523}, we separate the domain into two regions. For the solution of system (\ref{eq:equilibrium rho})-(\ref{eq:equilibrium V}) in the domain $r\in[0,r_{max}]$, we consider the central {\it core} region $r\in[0,r_{t}]$, $r_t<r_{max}$ with the density profile (\ref{eq:core density}), whereas in the subdomain $r\in[r_t,r_{max}]$ that we call the {\it tail} region, there will be a density profile resulting from a numerical integration. We call $r_t$ the transition radius between core and tail. Since the core density is prescribed by Eq. (\ref{eq:core density}), one only has to solve the system (\ref{eq:equilibrium rho})-(\ref{eq:equilibrium V}) in the {\it tail} region $r\in[r_t,r_{max}]$, which written as a first order system reads:

\begin{eqnarray}
\rho'_{tail} & = & u_{tail}, \label{eq:rhotail}\\
u_{tail}' & = & 4(V_{tail} + \frac{1}{2}v_{tail}^2 - V_0)\rho_{tail} \\
& &-\dfrac{2u_{tail}}{r} + \dfrac{u_{tail}^2}{2\rho_{tail}}, \label{eq:utail}\\
V'_{tail} & = & \dfrac{M_{tail}}{r^2}, \label{eq:Vtail}\\
M'_{tail} & = & r^2 \rho_{tail}, \label{eq:Mtail}\\
\bar{S}_{tail}' & = & v_{tail}, \label{eq:Stail}
\end{eqnarray}

\noindent where ${}^\prime := \frac{d}{dr}$ and $v_{tail} = \frac{A}{r^2 \rho_{tail}}$ is the velocity at the tail region, with $A$ constant, as suggested by the condition in Eq. (\ref{eq:equilibrium rho}). For the integration we impose the following boundary conditions at the two ends:

\begin{eqnarray*}
\rho_{tail}(r_t) & = & \rho_{core}(r_t), \\
u_{tail}(r_t) & = & u_{core}(r_t), \\
V_{tail}(r_t) & = & V_{core}(r_t), \\
M_{tail}(r_t) & = & M_{core}(r_t), \\
\bar{S}_{tail}(r_t) & = & 0, \\
M_{tail}(r_{max}) & = & M_{max}, 
\end{eqnarray*}

\noindent where $V_{core}$ and $M_{core}$ correspond to the solution of Poisson equation (\ref{eq:equilibrium V}) with  density (\ref{eq:core density}). Notice that the spatial part of the phase $\bar{S}$ is continuous at $r_t$, but it is not required to be differentiable, this implies that $v$ in general has a discontinuity at $r_t$. Moreover, the subsystem (\ref{eq:rhotail})-(\ref{eq:Stail}) is invariant under the change of sign in the velocity $v$ except by equation (\ref{eq:Stail}), this implies that two solutions can be constructed that satisfy the boundary conditions, one with inward and another one with outward radial velocity. 

We solve these equations using the shooting method with a fourth-order Runge-Kutta integrator that searches for the value of $A$, starting with an initial guess until the condition $M_{tail}(r_{max})=M_{max}$ is fulfilled within a tolerance. The results are the radial density profile at the tail, its radial velocity field and the gravitational potential.

Once the system (\ref{eq:rhotail})-(\ref{eq:Stail}) has been solved, it is possible to write down the wave function in the whole core-tail (CT) domain as follows:

\begin{equation}
\Psi_{CT}(t,r) \approx \left\lbrace
\begin{matrix}
\sqrt{\rho_{core}(r)}e^{-V_0 {\rm i}t} & r<r_t \\
&\\
\sqrt{\rho_{tail}(r)}e^{-V_0 {\rm i}t + {\rm i}\bar{S}_{tail}(r)} & r\geq r_t. \\
\end{matrix}
\right. \label{eq:MadelungInverse}
\end{equation}

\noindent Some illustrative examples used as workhorses are the following. If the transition radius is $r_t = 3.5 r_c \approx 4.58$ in units of the system of equations, a value consistent with those in \cite{Mocz:2017wlg}, assuming three values of the total mass to be $M_{max} = 10$, $20$, $40$,  and using the domain with $r_{max}=100$, the shooting parameter converges to the values $A \approx \pm 8.61\times10^{-2}$, $\pm 1.51\times10^{-1}$ and $\pm2.08\times10^{-1}$ respectively.

 In order to have an idea of what these three solutions represent physically, we show a particular example with units. Coordinates and mass are converted from dimensionless units into physical units according to the rule:

\begin{eqnarray}
r & \to & r_S r, \nonumber \\ 
t & \to & t_S t, \\
M & \to & M_S M, \nonumber 
\end{eqnarray}

\noindent where $t_S = \frac{m_B r_S^2}{\hbar}$ and $M_S = \frac{\hbar^2}{4\pi G m_B^2 r_S}$ are the scale factors for time and mass, respectively. These scale factors are expressed in terms of $r_S$, which is an arbitrarily chosen length scale factor, $m_B$ represents the boson mass, $\hbar$ is Planck's constant and $G$ Newton's gravitational constant. Other scale factors can be derived from these ones, such as the density scale factor $\rho_S = M_S/r_S^3$ and the velocity scale factor $v_S = r_S/t_S$. 

In order to use typical scales in a galaxy, let us consider a scale length of $r_S = 1.714$ kpc and a boson mass of $m_B = 10^{-23}$ eV/$c^2$. Using these values, the scale factors for time and mass are $t_S = 14.97$ Myr and $M_S = \frac{5}{4\pi}\times 10^{9}$ M${_\odot}$. Additionally, the scale factors for density and velocity are $\rho_S = 7.901\times10^{7}$ M${_\odot}$/kpc$^3$ and $v_S = 112.0$ km/s. Using these particular scale factors for the three example cases with $M_{max} = 10$, $20$, $40$, the transformed total masses correspond to $4\pi M_{\text{max}} = 5 \times 10^{10}$, $10^{11}$, and $2\times10^{11}$ M$_\odot$ and the core radius takes the value $r_c \approx 2.241$ kpc.

\begin{figure}
\includegraphics[width=8cm]{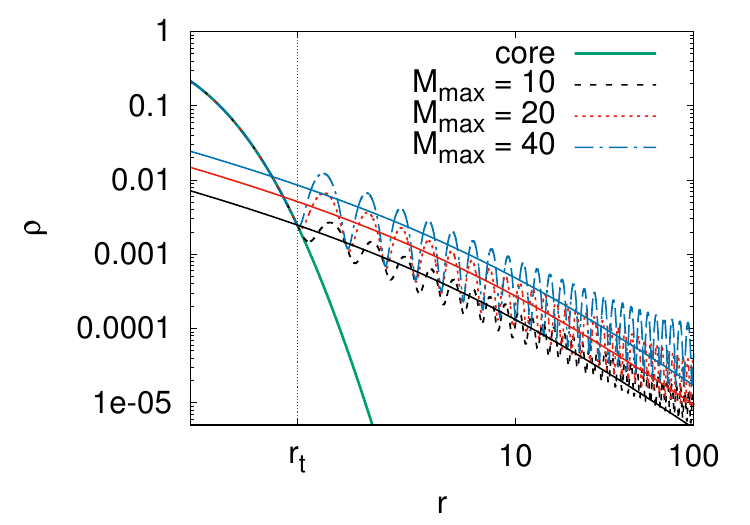}
\includegraphics[width=8cm]{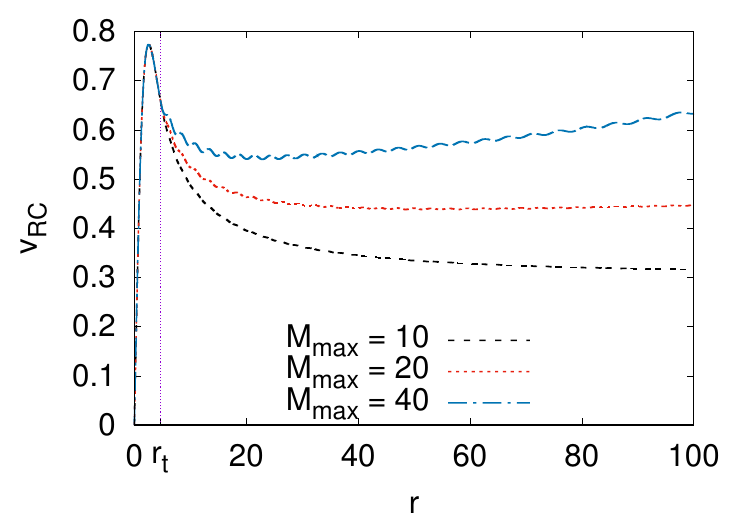}
\caption{(Top) Density profile of the solutions with total mass $M_{max} = 10$, $20$, $40$ and $r_{max}=100$. For 
$r < r_ t$ the profile is solitonic. Outside, for $r\ge r_t$, dotted lines correspond to the profile obtained for the equilibrium solutions, whereas the solid lines are the fittings of $\rho_{tail}$ using the NFW formula (\ref{eq:tail density}) that better fit the tail region. (Bottom) Rotation curves corresponding to the three solutions, where the trend of higher velocity for bigger mass can be noticed. }
\label{fig:solution equilibrium}
\end{figure}

 In Figure \ref{fig:solution equilibrium} we show the density profile of these three representative solutions. The density has the core profile (\ref{eq:core density}) for $r < r_t$, whereas for $r \ge r_t$ the dotted lines correspond to the solution in the {\it tail} zone of Eqs. (\ref{eq:rhotail})-(\ref{eq:Stail}). The density profile shows the typical oscillations of excited eigenmodes with $\ell =0$ explored in \cite{YavetzLiHui2022}. The resulting density profile admits a fitting with the NFW function:
 
\begin{equation}
\rho_{tail}(r) = \dfrac{\rho_{0,NFW}}{\frac{r}{R_s}\left(1+\frac{r}{R_s}\right)^2},
\label{eq:tail density}
\end{equation}

\noindent where $\rho_{0, core}$ and $R_s$ are free parameters that are specified individually for each halo, in our case for each core-tail solution. In the three illustrative cases, the fitting values in dimensionless units are $\rho_{0, tail} = 3.0\times10^{-2}, 4.3\times10^{-2}, 5.3\times10^{-2}$ and $R_s = 4.387, 4.399, 4.548$ for $M_{max} = 10$, $20$, and $40$ respectively.

In physical units, for the three specific examples, the parameters would take the values $\rho_{0, core}= 2.37\times 10^{6}, 3.40\times10^{6}, 4.19$ M$_\odot/$ kpc$^3$ and $R_s \approx 7.519, 7.540, 7.795$ kpc, obtained multiplying by the appropriate factors $\rho_S$ and $r_S$. After this parenthesis that illustrate how our core-halo solutions can be associated to galactic scales, we continue our discussion in dimensionless units with the three workhorse solutions illustrated here.

Notice that the matter in the tail region has a finite velocity given by $v_{tail}=\frac{A}{r^2 \rho_{tail}}$. The resulting solution in the whole domain is thus a stationary core surrounded by a tail with a shelled structure (see $\rho$ in Fig. \ref{fig:solution equilibrium}). Unlike in \cite{YavetzLiHui2022,PhysRevD.97.103523} the target of the solution is not the NFW profile, but $M_{max}$. This is a convenient aspect, because automatically we have a core-halo mass ratio as a possible target. Concerning the dependence of the solution on $r_{max}$, since the tail, with its oscillations never becomes zero, then the mass contained in a volume within spheres of different $r_{max}$ will be different. We verified that as long as the mean density is the same for different $r_{max}$, the solution is the same.
Additional to the density in Figure \ref{fig:solution equilibrium} we also show the Rotation Curve (RC) of the solutions constructed with the formula

\begin{equation}
v_{RC}(r) = \left\lbrace
\begin{matrix}
\sqrt{M_{core}(r)/r} & r<r_t, \\
&\\
\sqrt{M_{tail}(r)/r} & r\geq r_t .\\
\end{matrix}
\right.\label{eq:rc}
\end{equation}

\noindent Notice the trend of bigger total mass $M_{max}$ for higher rotation velocities. These three cases exemplify the effect of $M_{max}$ on the density profile and therefore on the rotation curve.

\begin{figure}
\includegraphics[width=8cm]{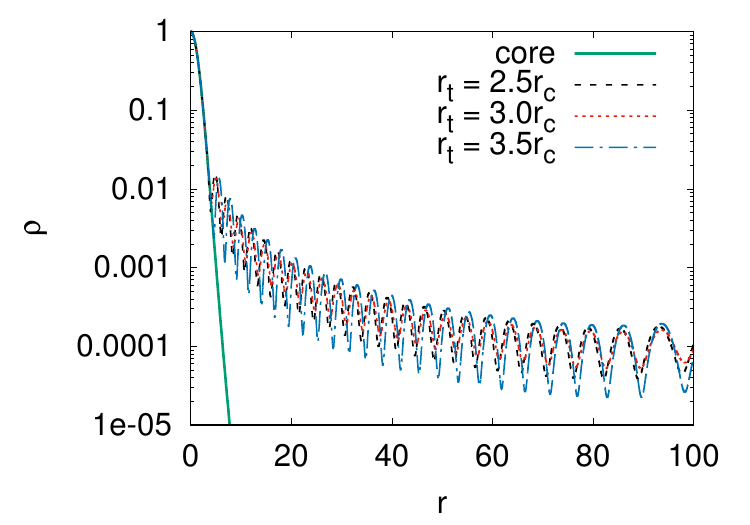}
\includegraphics[width=8cm]{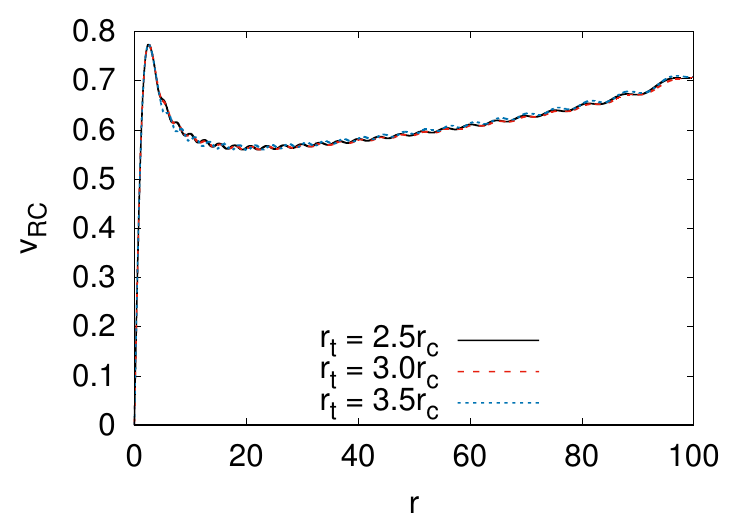}
\caption{(Top) Density profile of the solutions with $r_{t} = 2.5r_c$, $3.0r_c$ and $3.5r_c$ for the mass $M_{max}=50$.
(Bottom) Rotation curves of the three solutions. This result indicates that $r_t$ does not affect the tail profile as long as the total mass is the same.}
\label{fig:solution equilibrium2}
\end{figure}

We now look into the effects of the transition radius $r_t$. In Figure \ref{fig:solution equilibrium2} we show the results of three solutions with $r_t = 2.5r_c$, $3.0r_c$ and $3.5r_c$, an admisible range according \cite{Tula2018,Chan_2022}, and slightly bigger than the 2$r_c$ used in \cite{YavetzLiHui2022}, while this time we fix the target total mass to $M_{max}=50$ ($2.5\times 10^{11}{\rm M}_{\odot}$ in the units conversion example above). The core density is basically unaffected, whereas the tail density is different in the shell structure. Since the difference in the tail density profile is small, rotation curves are also very similar for the three values of the transition radius. 
The results of Figures \ref{fig:solution equilibrium} and \ref{fig:solution equilibrium2} imply that the rotation curves only depend significantly on the target total mass $M_{max}$ of the system and not on the transition radius $r_t$ that  separates core from tail.

\begin{figure}
\includegraphics[width=8cm]{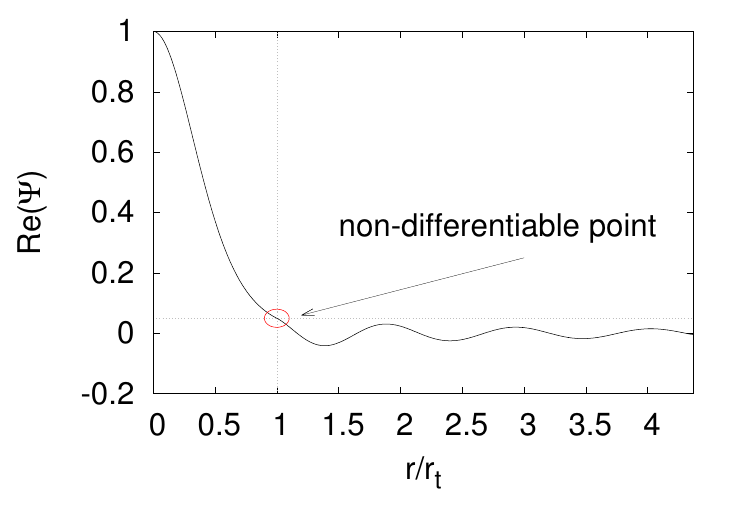}
\includegraphics[width=8cm]{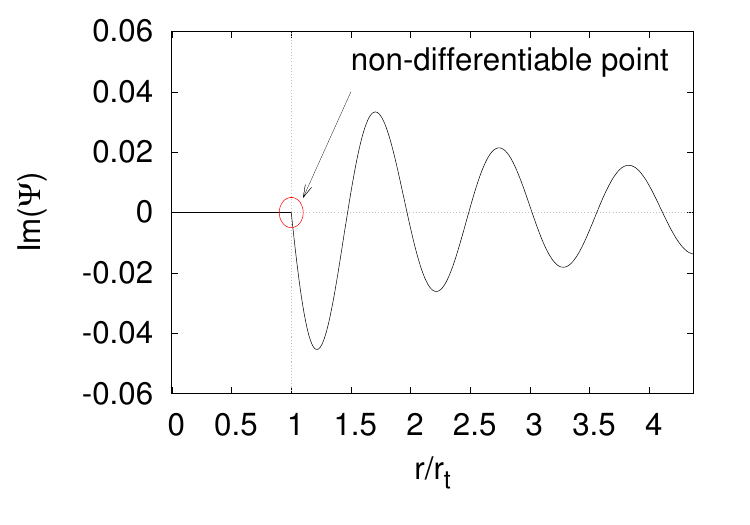}
\caption{At the top/bottom we show the real/imaginary part of the wave function. Notice the non-differentiability at $r_t = 3.5 r_c$ for the target mass $M_{tail}(100)=10$. This condition is the result of joining the core solution in the region $r<r_t$ with the tail solution in the region $r \ge r_t$.}
\label{fig:Repsi and ImPsi}
\end{figure}

{\it Discontinuity of $v$.} The phase $\bar{S}$ is continuous at $r_t$, however its derivative is not due to the condition  (\ref{eq:Stail}). We show the real and imaginary parts of the wave function of the solution in 
Figure \ref{fig:Repsi and ImPsi}, knowing that $\Psi = \sqrt{\rho} e^{iS}$, where $\rho$ and $S$ are the solutions of the system (\ref{eq:rhotail})-(\ref{eq:Stail}) for the case with $M_{tail}(100)=10$. It can be observed that there is a non-differentiability at the sphere of radius $r_t$ where core and tail are matched.
One would feel tempted to avoid this problem by manipulating the phase of the wave function of the solitonic core $\Psi(r,t)=e^{i\omega t}\psi(r)$, with $\omega$ the eigenvalue of the ground state equilibrium solution \cite{GuzmanUrena2004}, and one could manipulate the phase with an appropriate $t$ so that ${\rm Im}(\Psi)$ is non-zero with the appropriate continuity at $r=r_t$. However this would automatically imply that the phase $S$ inside the core is non-zero, and it would mean that the velocity there would be non-zero as well, which is inconsistent with stationarity of solitonic solutions \cite{AlvarezGuzmanMadelung}.

\begin{figure}
\includegraphics[width=8cm]{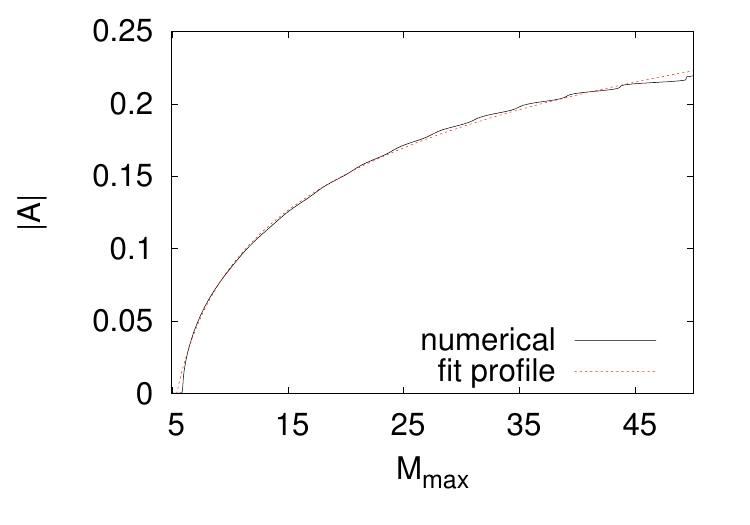}
\caption{Constant $A$ as function of $M_{max}$ obtained from the Shooting method and the fitting with a log function.}
\label{fig:A and rt0}
\end{figure}

Once we have seen that $M_{max}$ is the most influential parameter on the solutions, we collect information that depends on its value. We set $r_t = 3.5r_c$ and construct a bundle of solutions for $M_{max}\in[5,50]$. In Figure \ref{fig:A and rt0} we show the dependency of the constant $A$ as function of the boundary condition $M_{tail}(r_{max}=100) = M_{max}$, where the curve follows a logarithmic profile. 

%%%%%%%%%%%%%%%%%%%%%%%
\section{Evolution}
\label{sec:evolution}

As found in \cite{YavetzLiHui2022} for multimode solutions, there is an intrinsic tension between the assumption of stationarity and the result of the evolution. In or case, knowing that the velocity in the tail is non-zero, it is expected that the motion will affect the properties of the equilibrium solution. 
For the study of evolution we interpolate our wave function $\Psi_{CT}$ (\ref{eq:MadelungInverse}), into a 3D domain, where we solve the fully time-dependent equations in the Schr\"odinger frame (\ref{eq:Psi})-(\ref{eq:V}) using the code CAFE described in \cite{Alvarez_Rios_2022,periodicas}.

\subsection{Numerical methods for the evolution}

The evolution is carried out in a discrete domain described with Cartesian coordinates $D_d = \{(x_i,y_j,z_k)\in D=[x_{min},x_{max}]\times[y_{min},y_{max}]\times[z_{min},z_{max}] | x_i = x_{min} + i h, y_j = y_{min} + j h, z_k = z_{min} + k h\}$ where $i,j,k = 0,1,...,N-1$, $h=(x_{max}-x_{min})/(N-1)$ is the spatial resolution and $N$ is the number of  points along each spatial direction, whereas time is discretized with $t^n = n\Delta t$ where $\Delta t$ is the temporal resolution and $n=0,1,..,N_t$ with $N_t$ the number of evolution time steps. 
The system is evolved using the Crank-Nicolson time-average, which is written as

\begin{equation}
\left(1+\frac{1}{2}{\rm i}\Delta t \hat{H}^{n+1}\right)\Psi^{n+1} = \left(1-\frac{1}{2}{\rm i}\Delta t \hat{H}^{n}\right)\Psi^{n},
\label{eq:Crank-Nicolson} 
\end{equation}

\noindent where $\Psi^n = \Psi(t^n,\vec{x})$ and $\hat{H}^n=-\frac{1}{2}\nabla^2 + V^n$ is the Hamiltonian operator at time $t^n$. For the solution of this equation, we split kinetic and potential energies in a scheme called \textit{``kick-drift-kick"} step splitting:

\begin{equation}
\Psi^{n+\alpha} = \dfrac{1-\frac{1}{4}i\Delta t V^{n+1/2}}{1+\frac{1}{4}i\Delta t V^{n+1/2}}\Psi^n, 
\label{eq: evolve first step}
\end{equation}

\begin{equation}
\Psi^{n+\beta} = \mathcal{F}^{-1}\left[\dfrac{1-\frac{1}{4}i\Delta t k^2}{1+\frac{1}{4}i\Delta t k^2}\mathcal{F}\left(\Psi^{n+\alpha}\right)\right], 
\label{eq: evolve second step}
\end{equation}

\begin{equation}
\Psi^{n+1} = \dfrac{1-\frac{1}{4}i\Delta t V^{n+1/2}}{1+\frac{1}{4}i\Delta t V^{n+1/2}}\Psi^{n+\beta}, 
\label{eq: evolve third step}
\end{equation}

\noindent where the potential at time $t^n$ is obtained as

\begin{equation}
V^{n} = -\mathcal{F}^{-1}\left[\dfrac{\mathcal{F}(|\Psi^n|^2  - \expval{|\Psi^n|^2})}{k^2}\right],
\end{equation}

\noindent where $V^{n+1/2} = (V^{n+1} + V^n)/2$ and 

\begin{equation}
\mathcal{F}\left\lbrace \Psi^n \right\rbrace(\vec{k}) := \int_{\textup{all space}} \Psi^n(\vec{x}) e^{i\vec{k}\cdot\vec{x}} d^3x,
\end{equation}

\noindent is the Fourier transform, $\mathcal{F}^{-1}$ its inverse, $\vec{k}$ the wave number vector and $k=|\vec{k}|$ its magnitude. The calculation of the Fourier Transform and its inverse uses the Fast Fourier Transform ($fft$) and its inverse ($ifft$), suitable for periodic boundary conditions that are imposed on functions $\Psi$ and $V$. We use a periodic domain because it prevents matter from escaping to infinity as discussed in \cite{periodicas} and used in \cite{YavetzLiHui2022}.

% --------------------------------
\subsection{Diagnostics}

In order to monitor the evolution of the system we calculate the total mass defined as $M = \int_D |\Psi|^2d^3x$, in order to verify the behavior the of energy, we calculate the kinetic energy $K = -\frac{1}{2}\int_D \Psi^* \nabla^2\Psi d^3x$ and the potential energy $W = \frac{1}{2}\int_D V|\Psi|^2 d^3x$, which are useful to obtain the total energy $K+W$ and the virilization function $2K+W$.

{\it Configurations evolved.} We use the same three equilibrium solutions described above, with $M_{max}=10$, 20, 40 and transition radius at $r_t=3.5r_c$, whose density profiles are those in Figure \ref{fig:solution equilibrium}. The simulation is carried out in two spatial domain sizes, small $D=[-20,20]^3$ and big $D=[-40,40]^3$, using spatial resolutions $40/63$ and $80/127$, respectively. The temporal domain is $t\in[0,600]$ and time resolution is $\Delta t=0.1$ in all cases. We want to stress the importance of the domain size in this type of simulations, because its effects are barely discussed. In \cite{periodicas} we elaborate about the effects of domain size on simulations of various scenarios when using periodic boundary conditions, chief among them the fact that the matter outside a core redistributes differently in different domain sizes. This is why density and interference patterns developed in the small big domains are necessarily different. We use two box sizes precisely to demonstrate that average density and average RCs do not vary significantly.

In order to have an idea of the domain solution in physical units, for the boson mass and length scale used above for illustration, small and big  numerical domains are cubic boxes of side 68.56 and 137.12 kpc respectively. The time domain on the other hand translates 600 dimensional units into 8.984 Gyr.

\begin{figure}
\includegraphics[width=4.25cm]{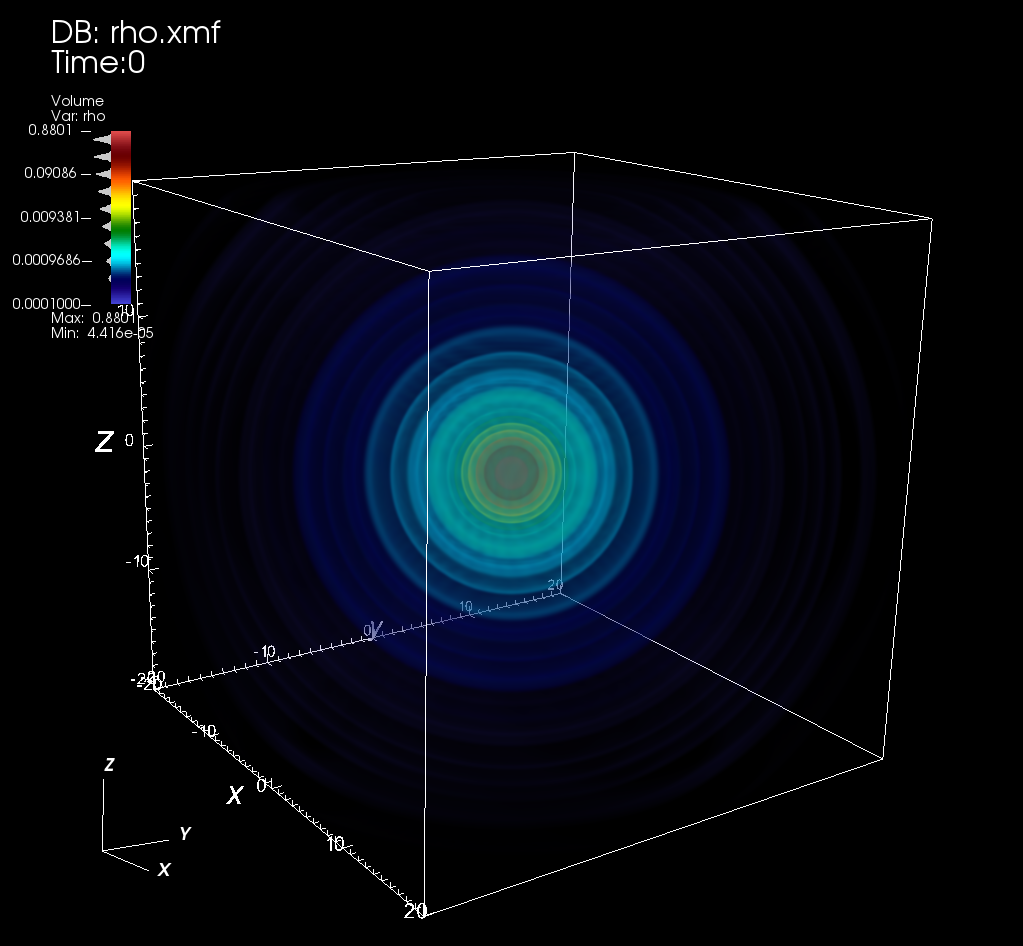}
\includegraphics[width=4.25cm]{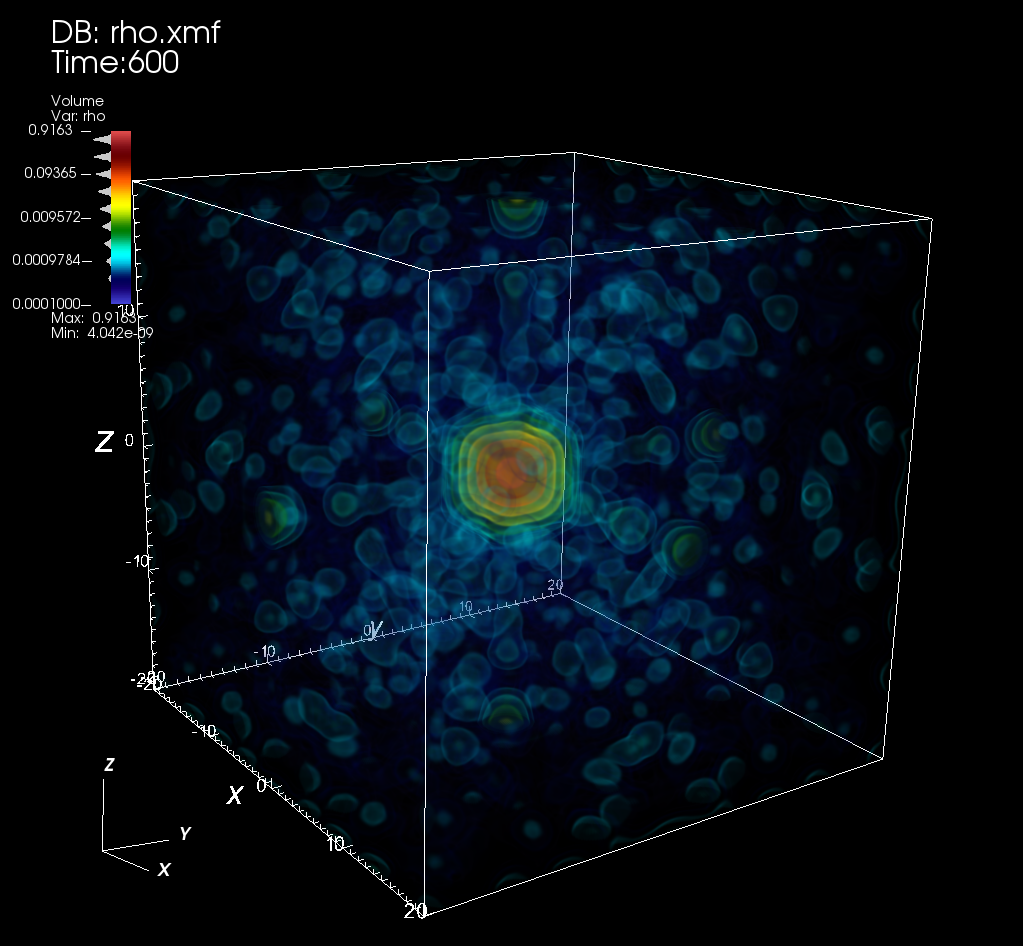}
\includegraphics[width=4.25cm]{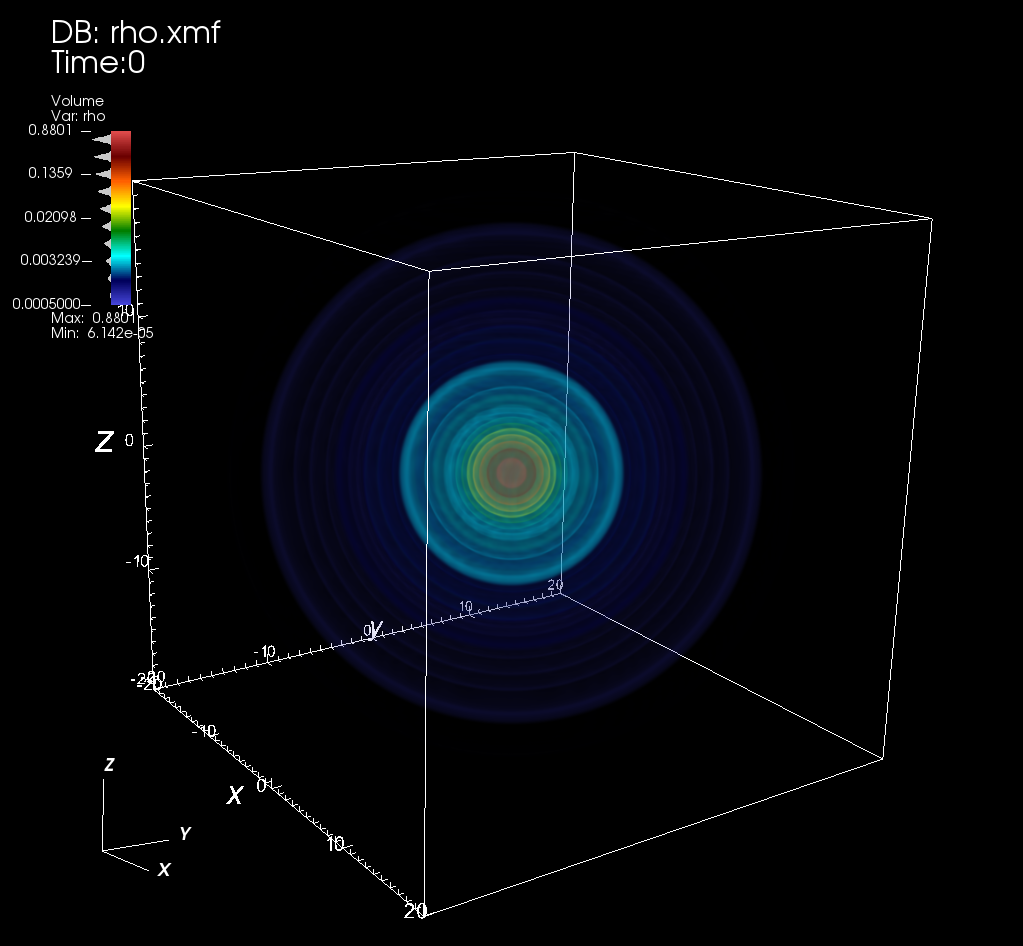}
\includegraphics[width=4.25cm]{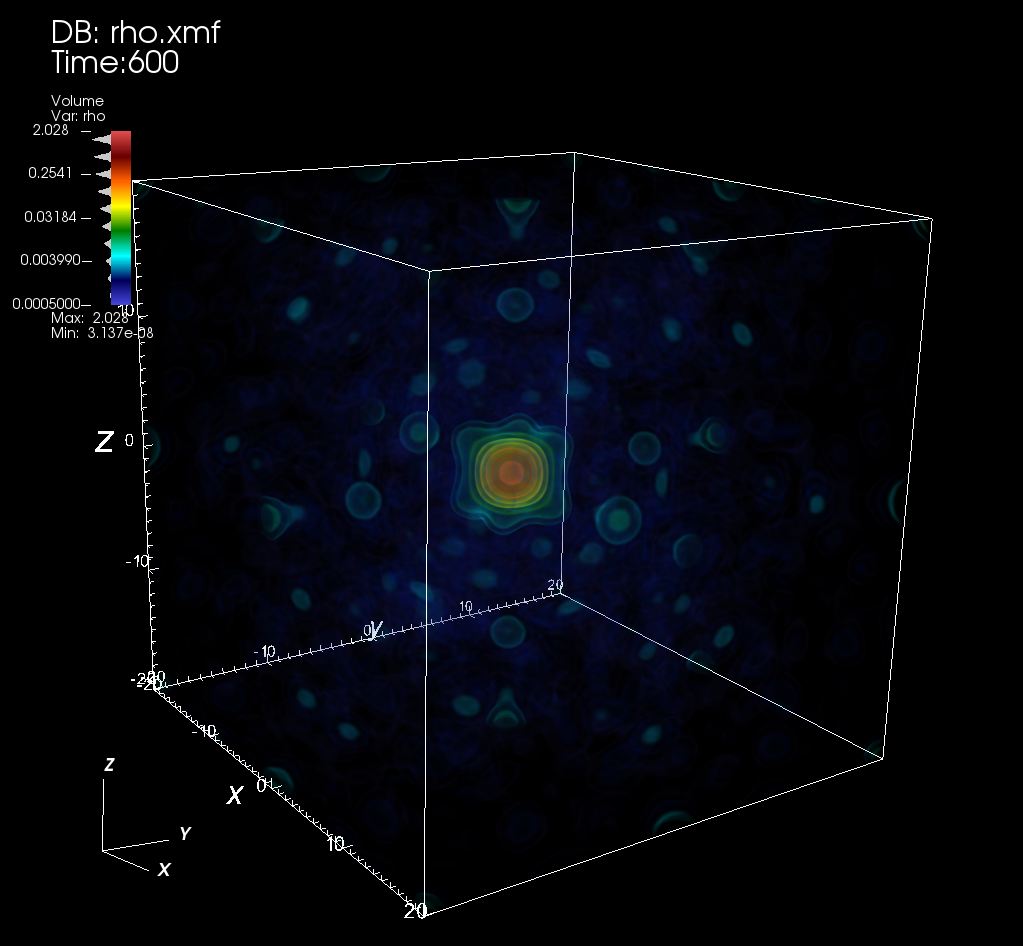}
\includegraphics[width=4.25cm]{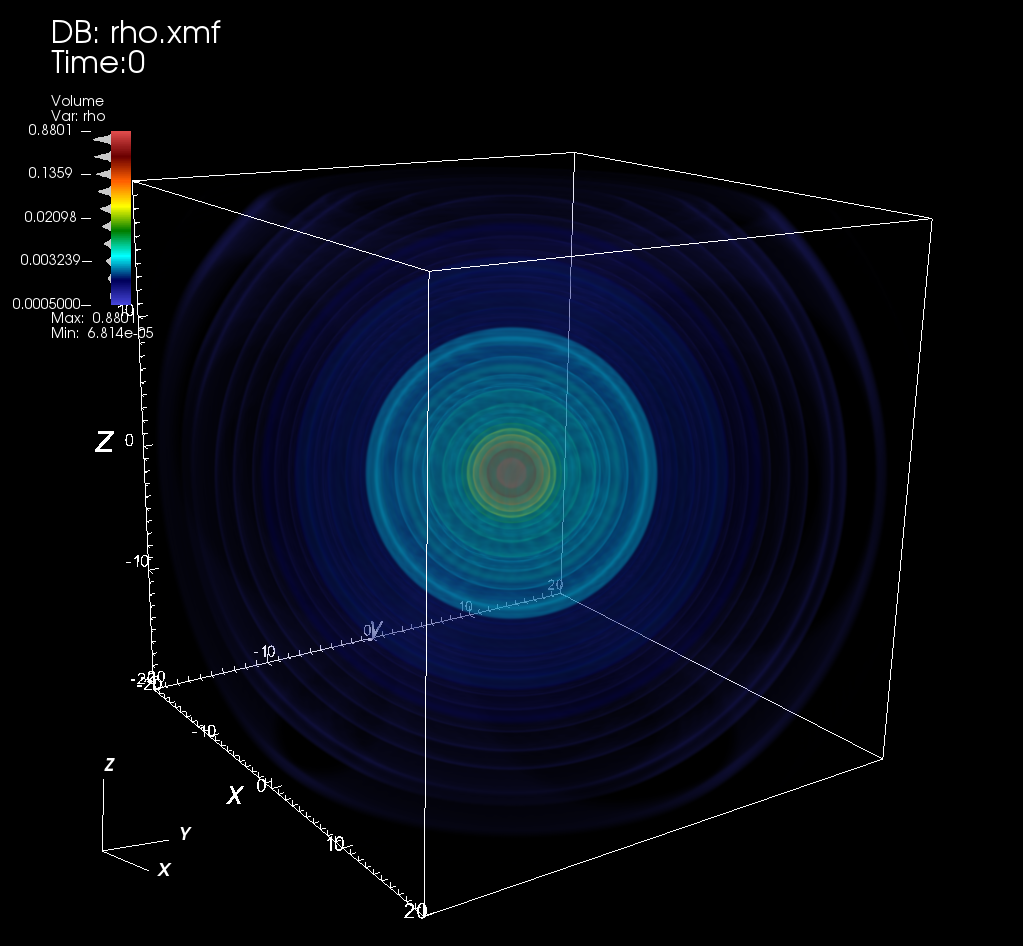}
\includegraphics[width=4.25cm]{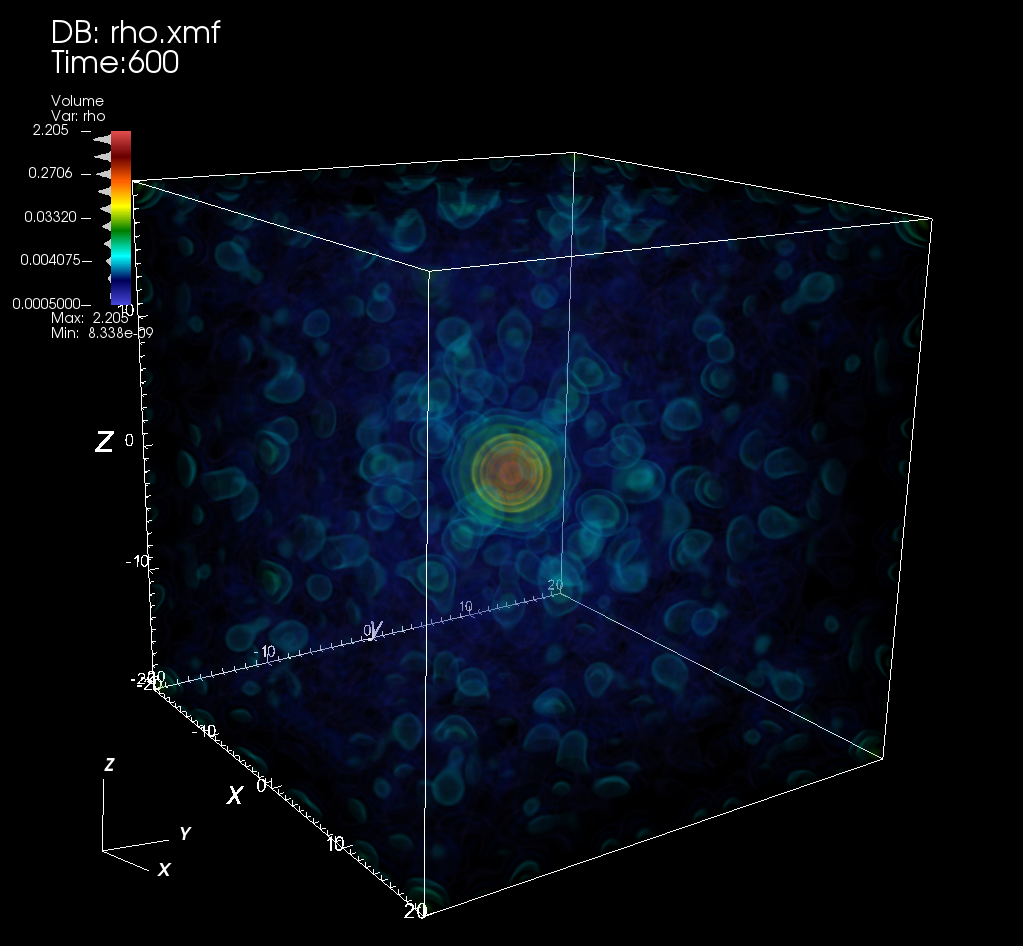}
\caption{Snapshots of the density $|\Psi|^2$ at times $t=0$ and 600 for the evolution of equilibrium configurations with boundary conditions $M_{max}=10$ (top), 20 (middle) and 40 (bottom) using transition radius $r_t=3.5 r_c$. The spatial domain is $D=[-20,20]^3$.}
\label{fig:snaps small}
\end{figure}

\begin{figure}
\includegraphics[width=4.25cm]{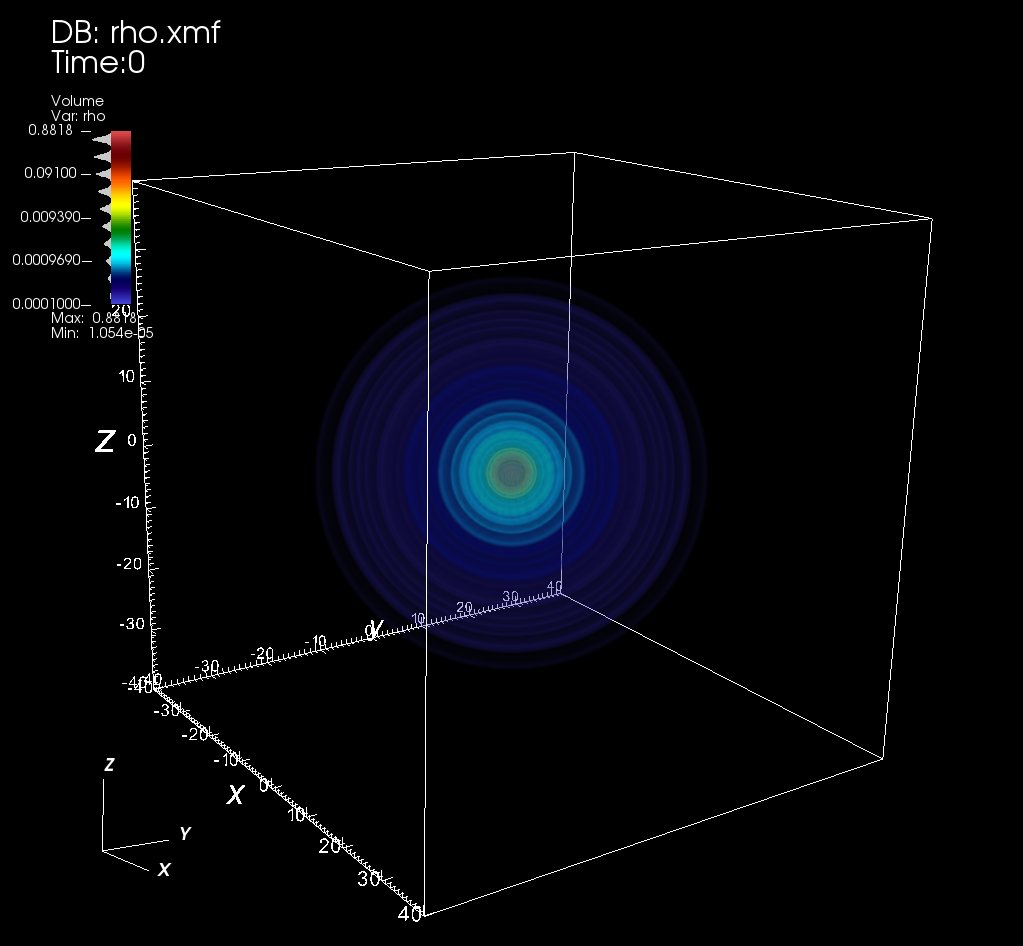}
\includegraphics[width=4.25cm]{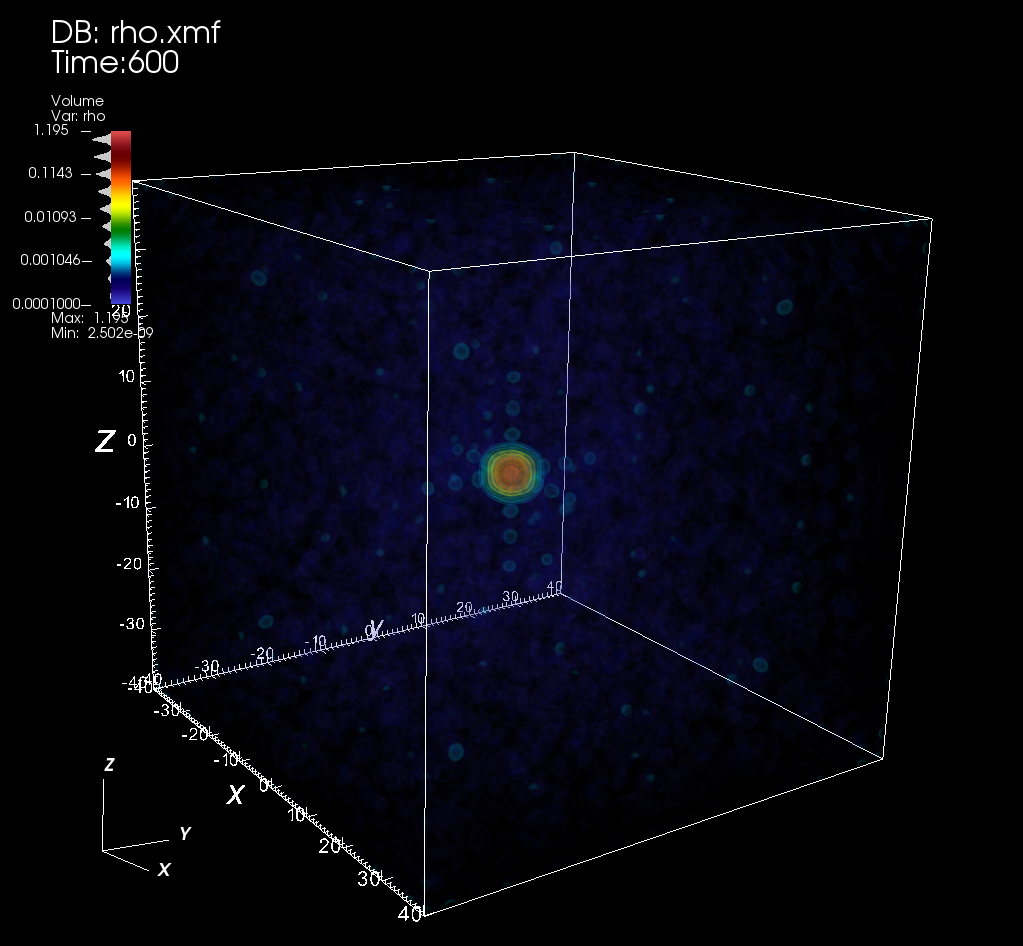}
\includegraphics[width=4.25cm]{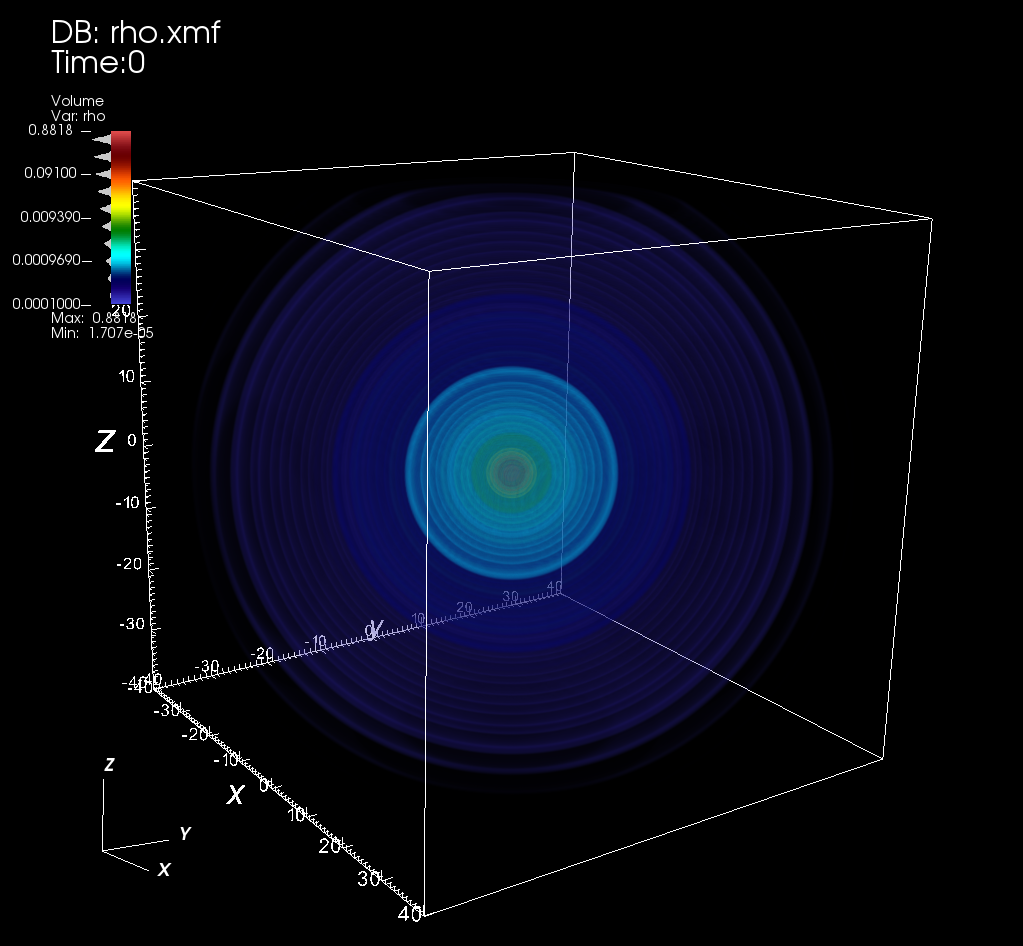}
\includegraphics[width=4.25cm]{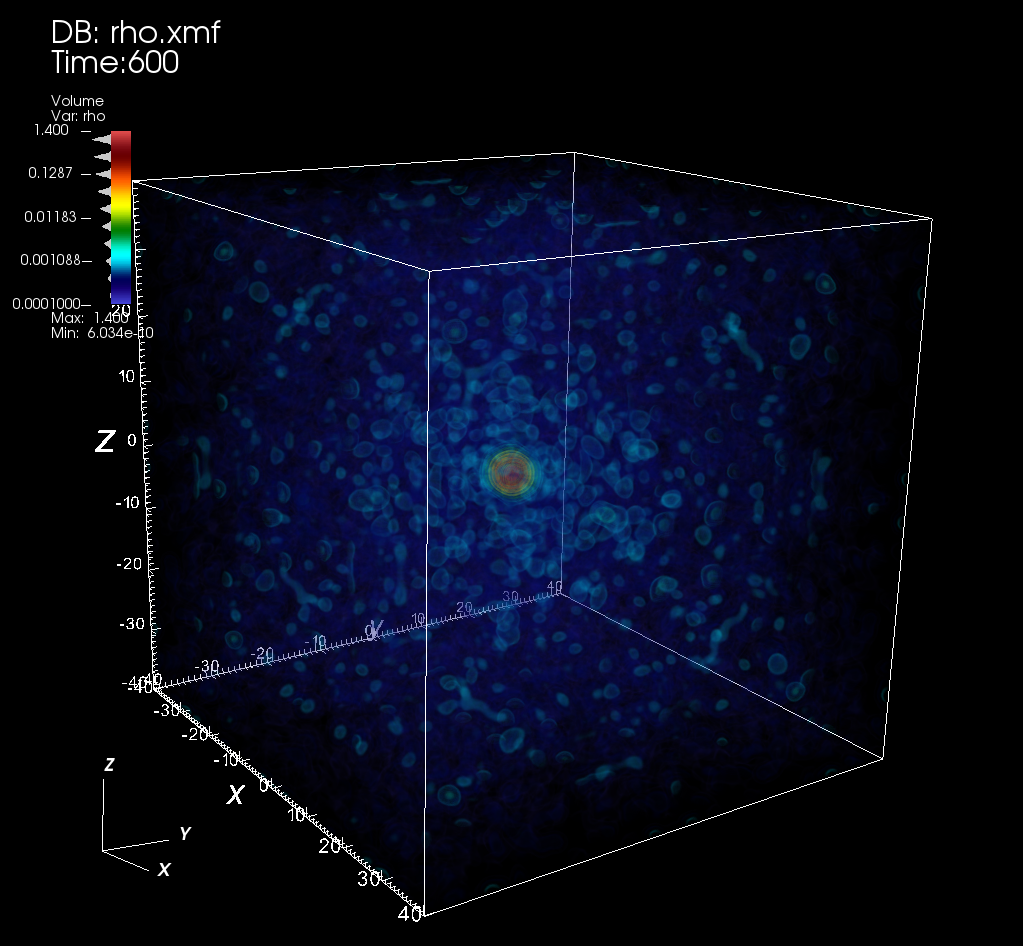}
\includegraphics[width=4.25cm]{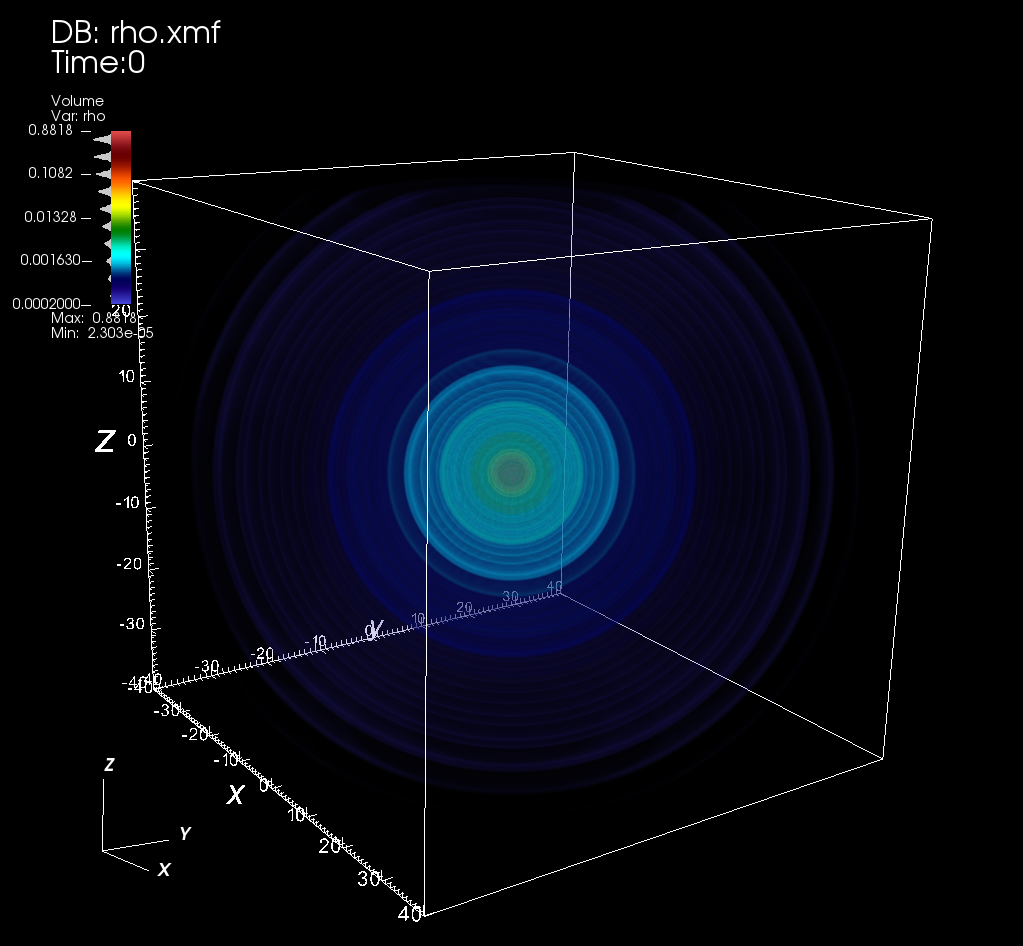}
\includegraphics[width=4.25cm]{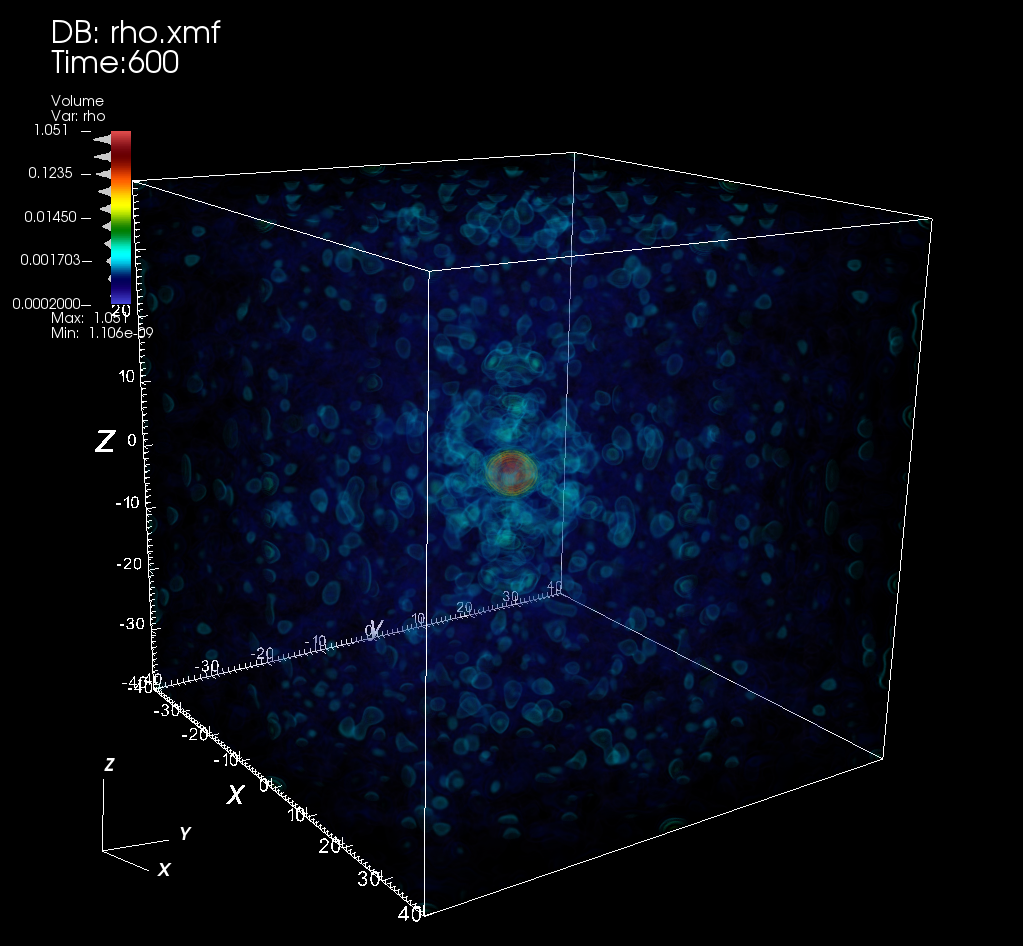}
\caption{Snapshots of the mass density at times $t=0$ and 600 for the evolution of equilibrium configurations with boundary conditions $M_{max}=10$ (top), 20 (middle) and 40 (bottom) are shown with transition radius of $r_t=3.5 r_c$. The spatial domain is $D=[-40,40]^3$.}
\label{fig:snaps big}
\end{figure}

In Figures \ref{fig:snaps small} and \ref{fig:snaps big} we show snapshots of $|\Psi|^2$ at initial time and $t=600$, for the evolution of the three solutions with $M_{max}=10$, 20, 40, illustrating the development of the system in the small and big domains. The dynamics is triggered by the initial radial velocity of the matter at the tail region that perturbs the core and redistributes itself due to the periodic boundary conditions, which produces the typical interference observed in simulations of structure formation simulations and collisions of multiple cores \cite{Schive:2014dra,Mocz:2017wlg,Veltmaat_2018,periodicas}.

\begin{figure}
\includegraphics[width=4.2cm]{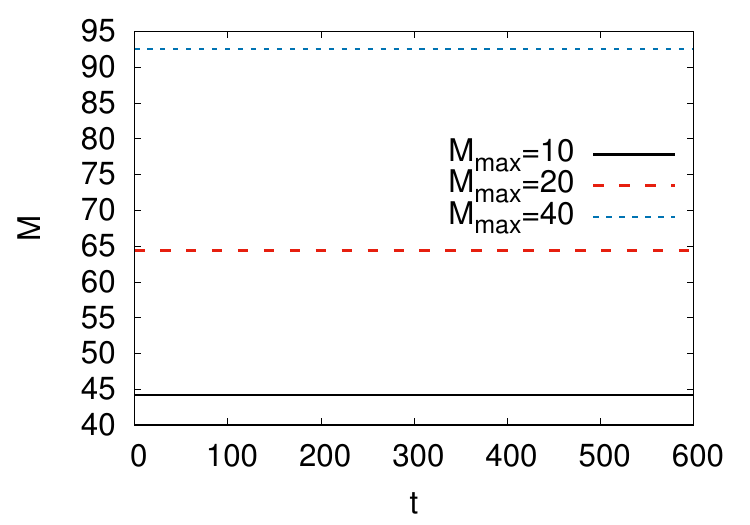}
\includegraphics[width=4.2cm]{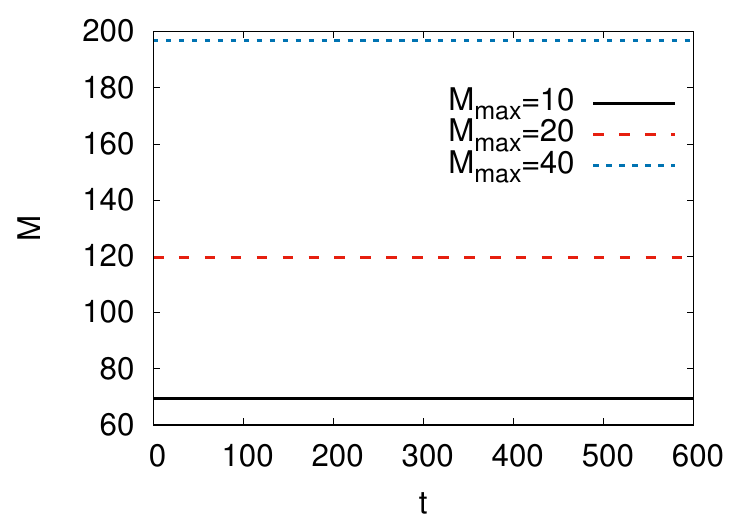}
\includegraphics[width=4.2cm]{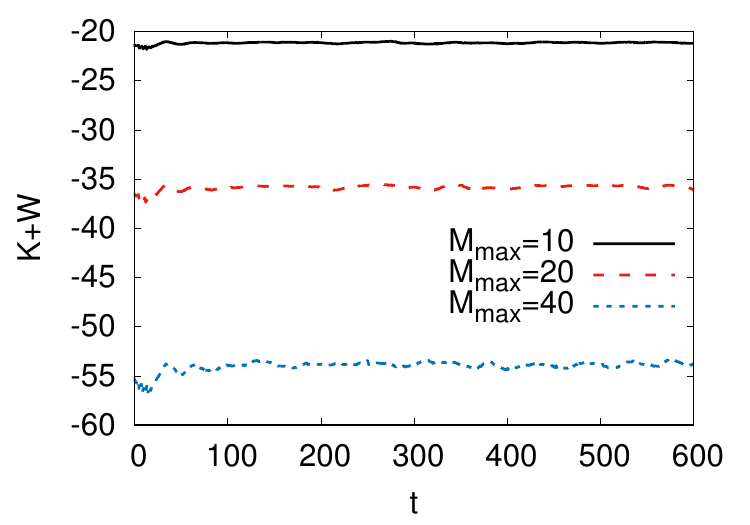}
\includegraphics[width=4.2cm]{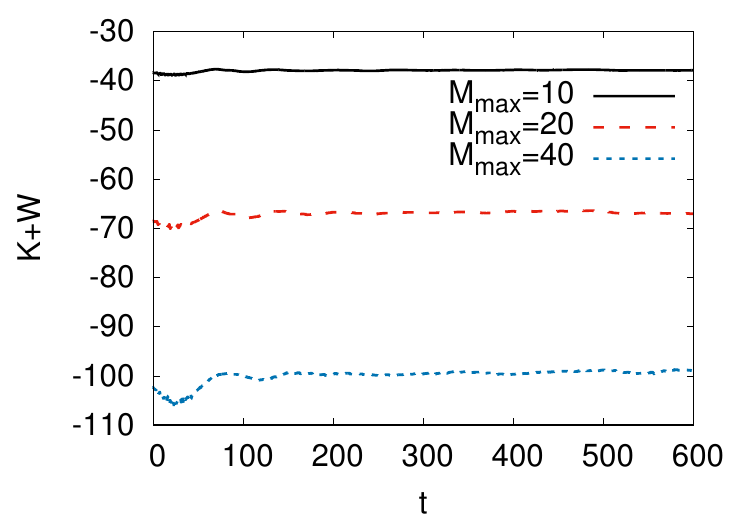}
\includegraphics[width=4.2cm]{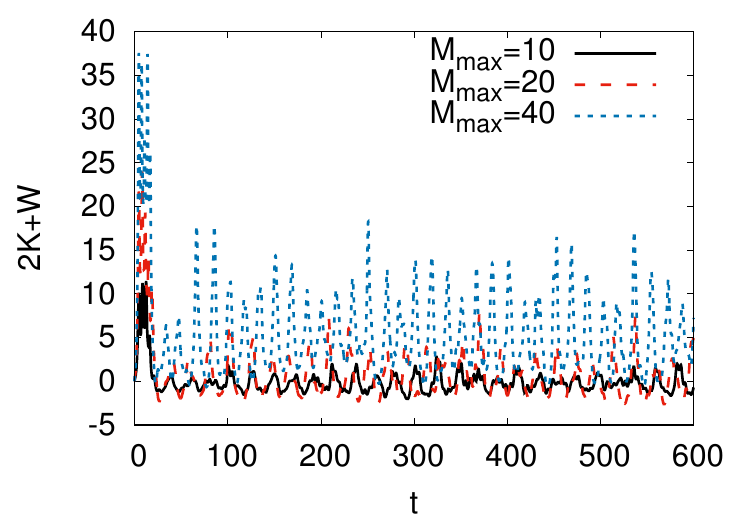}
\includegraphics[width=4.2cm]{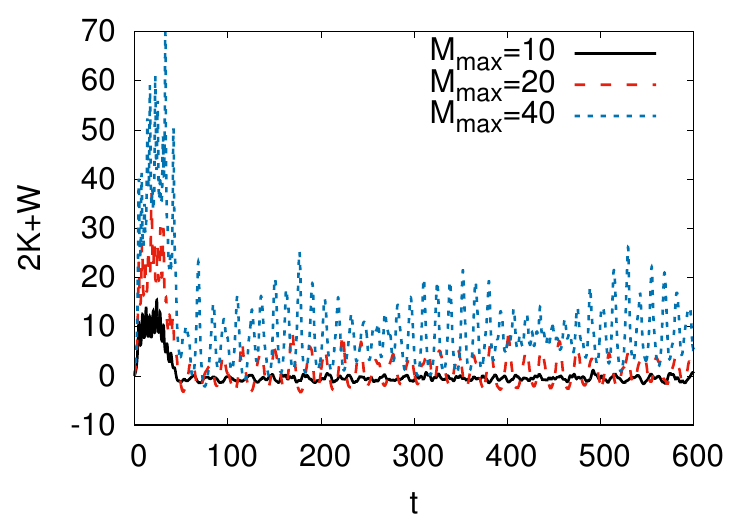}
\includegraphics[width=4.2cm]{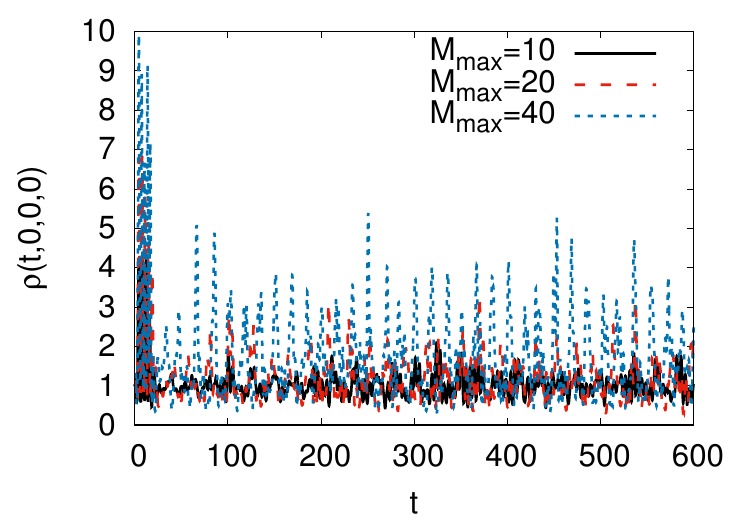}
\includegraphics[width=4.2cm]{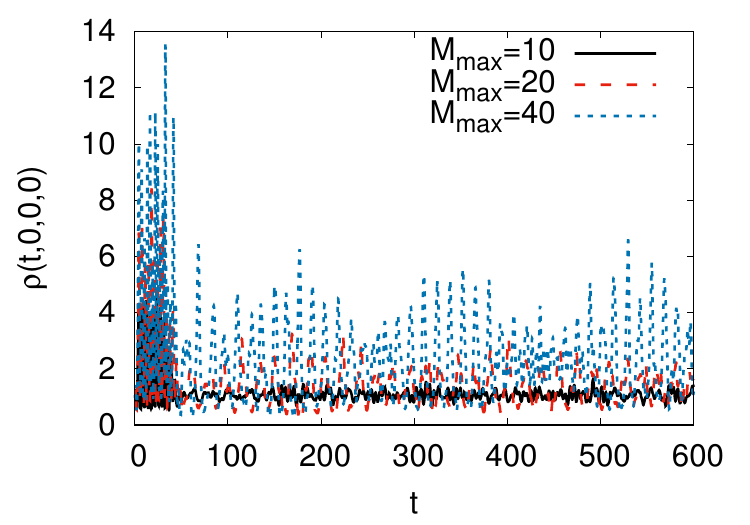}
\caption{Scalars monitored during the evolution in the small domain $D=[-20,20]^3$ (left column) and big domain $D=[-40,40]^3$ (right column). In the first row, the mass as function of time that shows preservation. In the second row, the total energy shows an initial transient and afterwards remains approximately constant. In the third row, the quantity $2K+W$ illustrates how the system is initially virialized, gets distorted during the initial transient and later on it oscillated near a virialized state. In the fourth row the central density shows that initially the configuration collapses, then expands  and approaches a value close to its initial value.}
\label{fig:diagnosis}
\end{figure}

In Figure \ref{fig:diagnosis} we present the evolution of some scalars as function of time, at the left/right we show diagnostics using the small/big domain. The first row shows evolution of the mass, which is conserved during the evolution. The second row displays the total energy as function of time, which after an initial transient the total energy remains approximately constant over time. The third row depicts the quantity $2K+W$, which is initially zero but then increases due to the initial transient, then it starts oscillating near zero, which indicates that the configuration tends to approach a virialized state again that we call the {\it relaxed} configuration. Finally, the fourth row shows the density at the origin as function of time, which after the initial transient oscillates regularly. The initial transient is due to the fact that we set the constant $A$ to be positive, thus the tail velocity points toward the core, which produces an incoming flow,  compresses the core initially and the evolution is followed by the expansion and stabilization around a virial state.
Notice that the bigger the value of $M_{max}$ the bigger the amplitude of oscillations of central density and $2K+W$. The reason is that for bigger $M_{max}$ the {\it tail} is more massive and therefore at initial time the momentum transferred to the core is bigger, which adds a bigger perturbation to the system.

% --------------------------------
\subsection{Equilibrium solution vs relaxed solution}

\begin{figure}
\includegraphics[width=4.2cm]{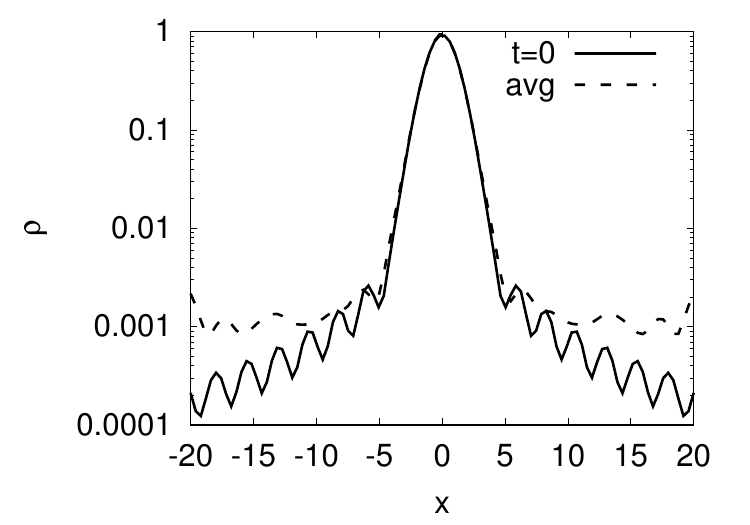}
\includegraphics[width=4.2cm]{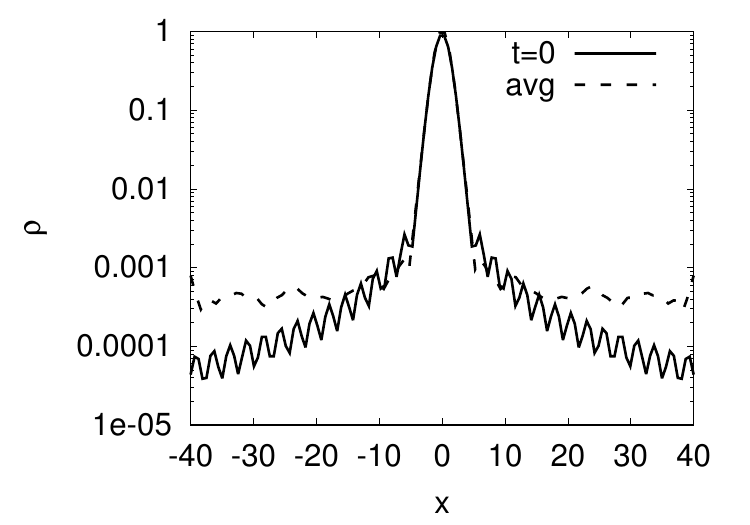}
\includegraphics[width=4.2cm]{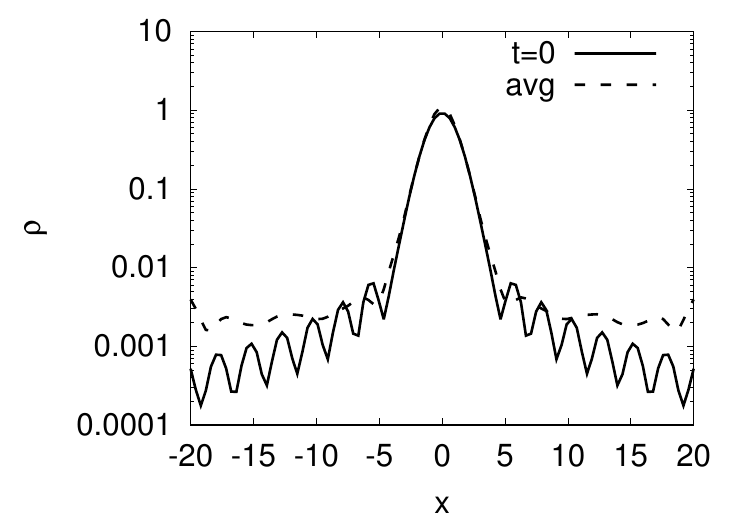}
\includegraphics[width=4.2cm]{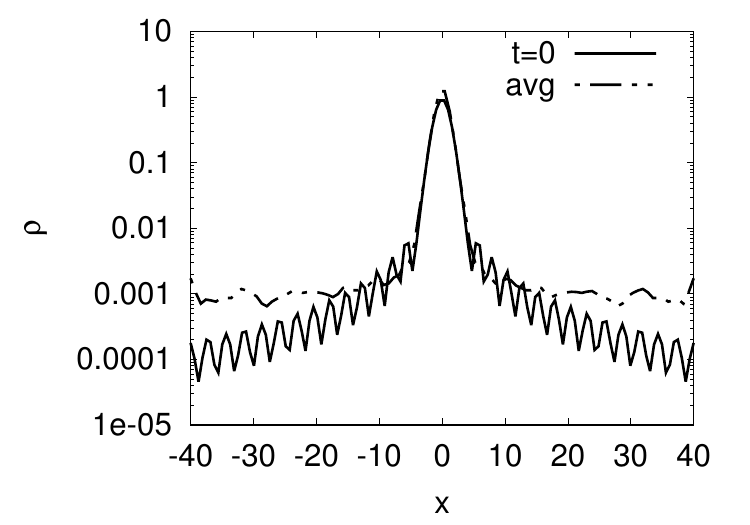}
\includegraphics[width=4.2cm]{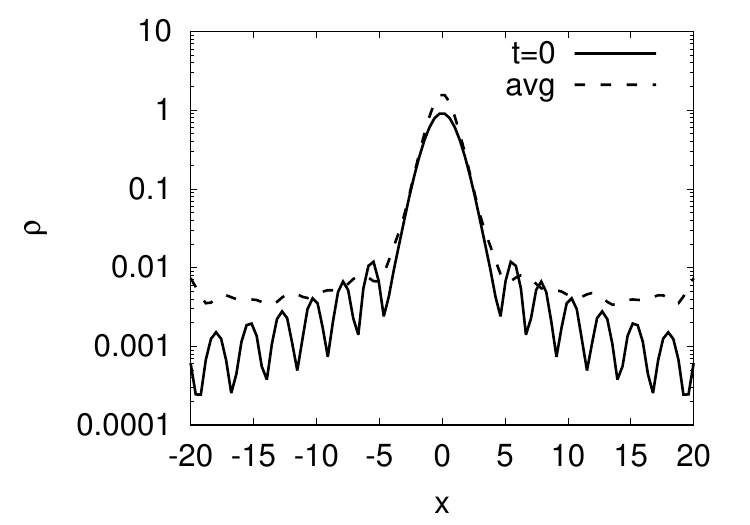}
\includegraphics[width=4.2cm]{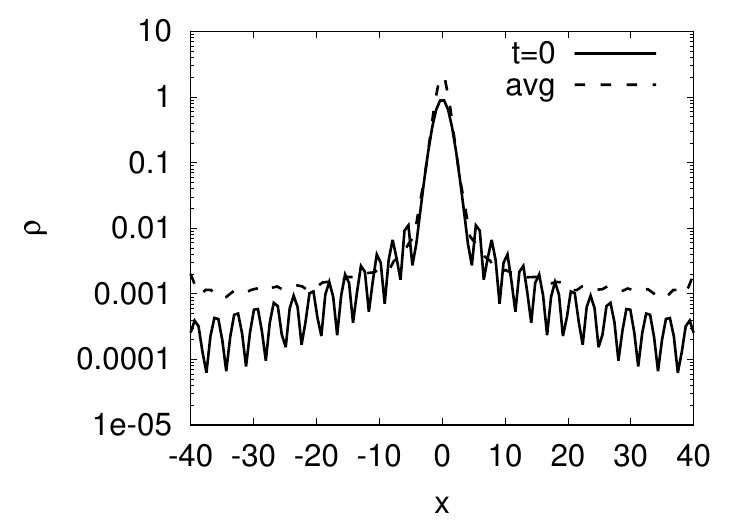}
\caption{In these plots, continuous/dashed lines correspond to the density of {\it equilibrium/relaxed} configurations. The left column shows the density when using the small domain $D=[-20,20]^3$, and the right one displays the results  using the big domain $D=[-40,40]^3$. The first, second, and third rows correspond to the cases with $M_{max} = 10$, 20, and 40, respectively.}
\label{fig:rho}
\end{figure}

{\it Density}. We investigate the utility of equilibrium configurations by studying their evolution and assessing how much the matter distribution deviates from the initial profile. In Figure \ref{fig:rho} we show the density of the {\it equilibrium} configurations with continuous lines and the average of {\it relaxed} configurations with dashed lines. On the left column the results correspond to the small domain $D = [-20, 20]^3$, and on the right in the big domain $D = [-40, 40]^3$. The  three rows correspond to $M_{max}$ = 10, 20, and 40, respectively. The differences between the equilibrium and relaxed profiles are greater for larger $M_{max}$. This can be explained because according to Figure \ref{fig:A and rt0}, the shooting target constant $A$ is a growing function of $M_{max}$, and hence $v_{tail}$ increases as $A$ grows. Moreover, the bigger the $M_{max}$ the bigger the mass of the tail and therefore the momentum carried by the tail grows with total mass and stronger the kick. In the end, the density and rotation curves of the relaxed configuration deviate more from the equilibrium solution for larger $M_{max}$.

\begin{figure}
\includegraphics[width=4.2cm]{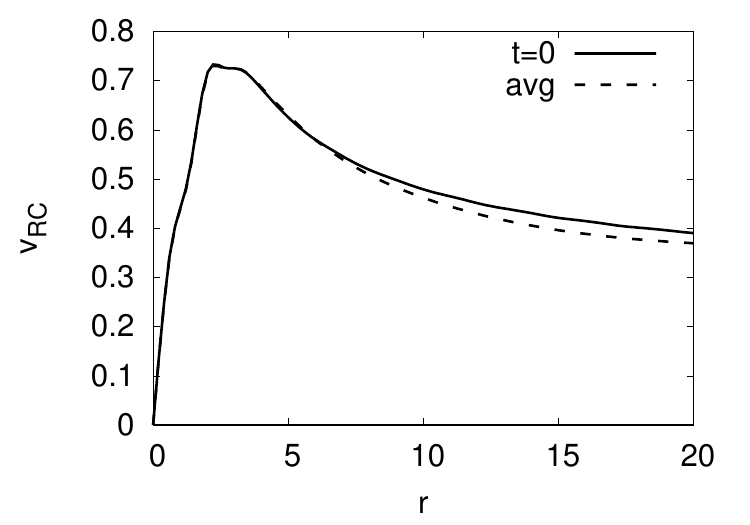}
\includegraphics[width=4.2cm]{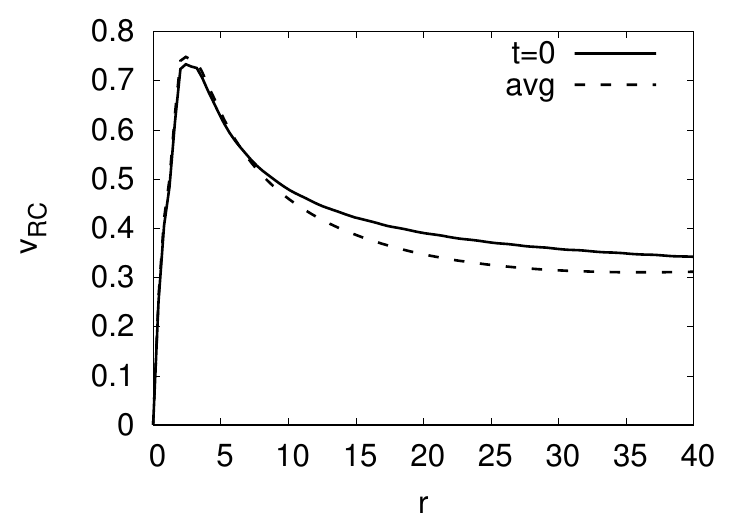}
\includegraphics[width=4.2cm]{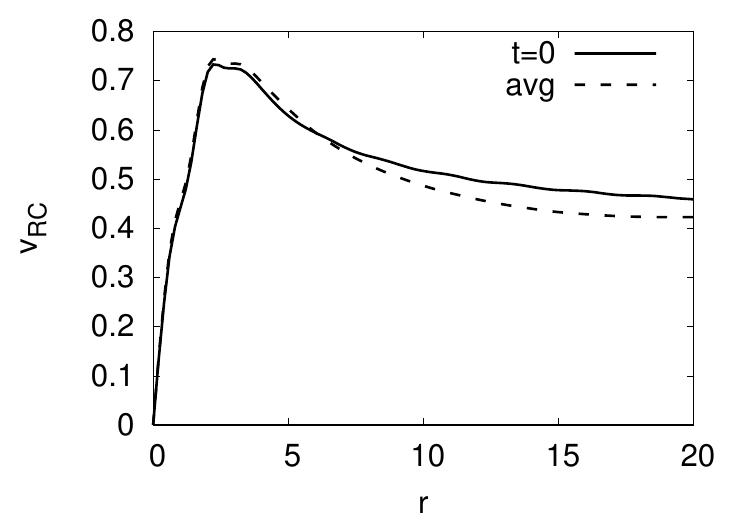}
\includegraphics[width=4.2cm]{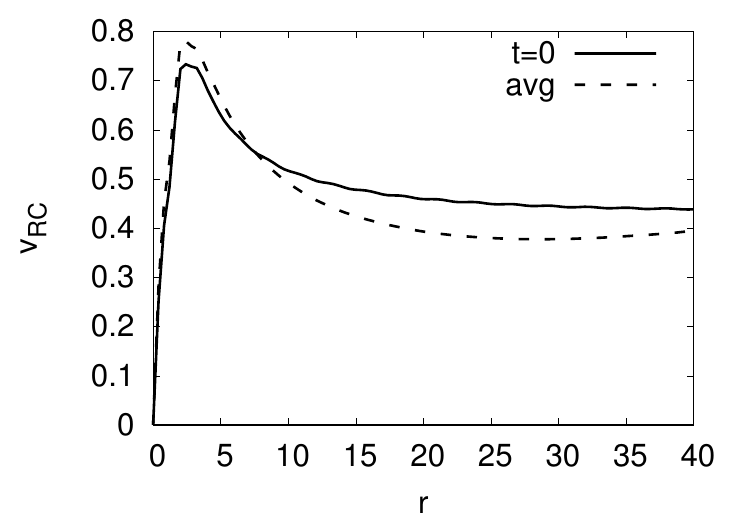}
\includegraphics[width=4.2cm]{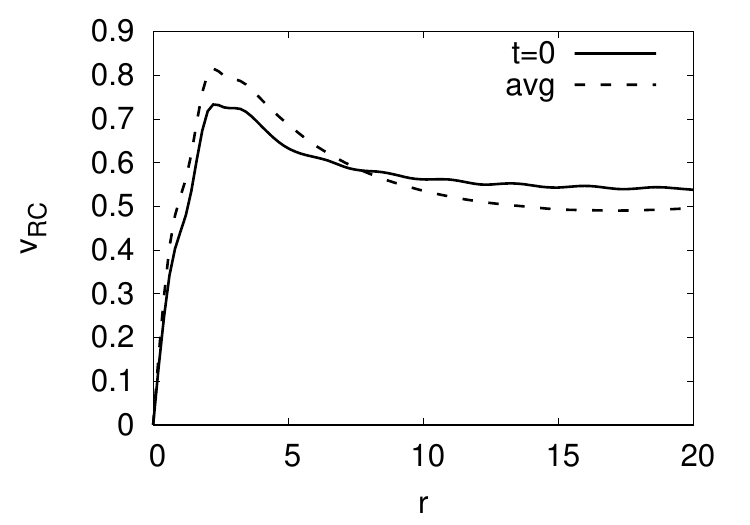}
\includegraphics[width=4.2cm]{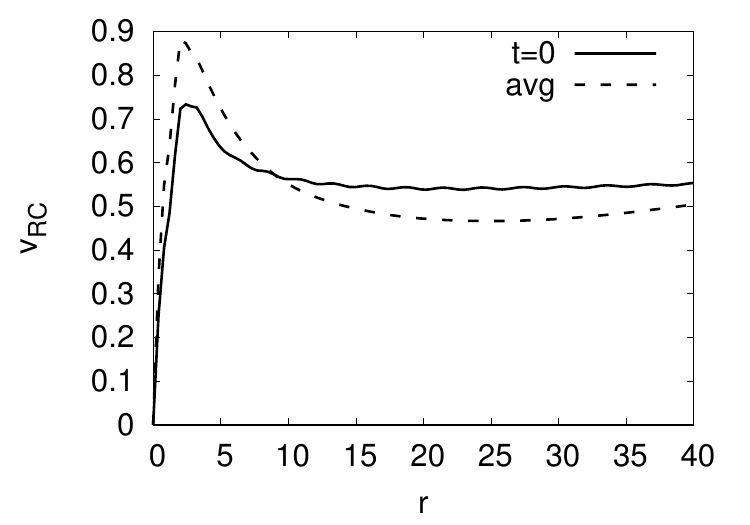}
\includegraphics[width=4.2cm]{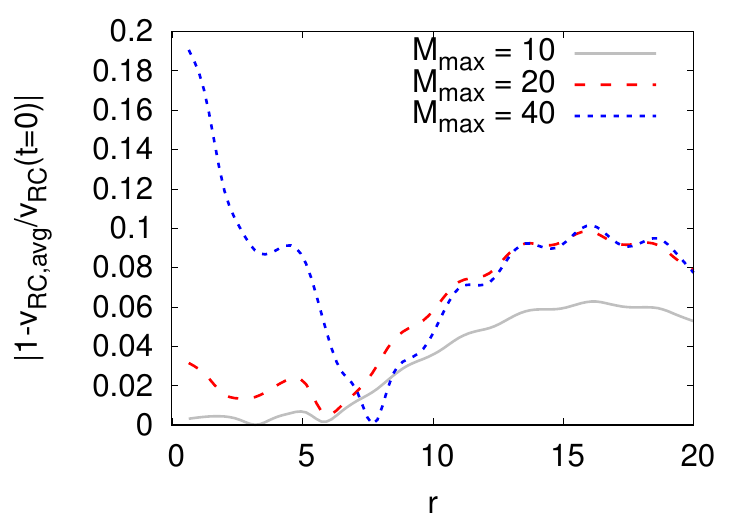}
\includegraphics[width=4.2cm]{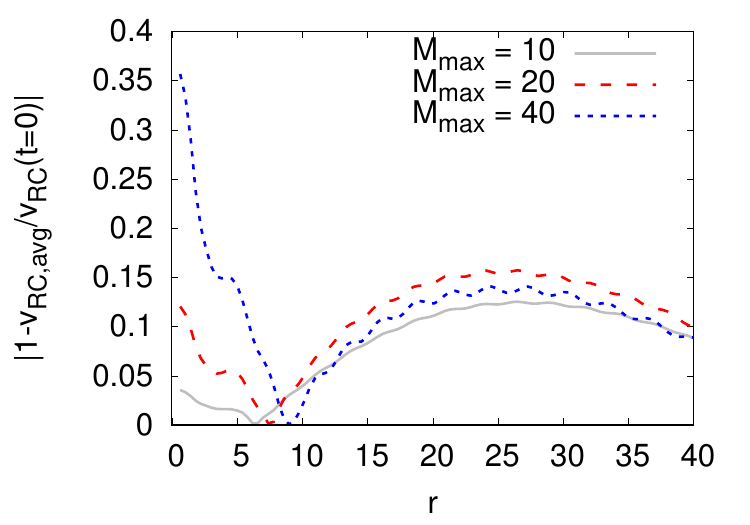}
\caption{In these plots, continuous/dashed lines correspond to the RCs of {\it equilibrium/relaxed} configurations. The left column shows the density when using the small domain $D=[-20,20]^3$, and the right one displays the results  using the big domain $D=[-40,40]^3$. The first, second, and third rows correspond to the cases with $M_{max} = 10$, 20, and 40, respectively. At the bottom row we plot the relative difference between the RC for the {\it equilibrium} and {\it relaxed} configurations. In physical units, for the example worked out in the text, one has to multiply $v_{RC}$ by 112km/s and $r$ by 1.714 kpc.}
\label{fig:RC}
\end{figure}

{\it Rotation Curves.} We also analyze RCs of {\it equilibrium} and {\it relaxed} configurations. Figure \ref{fig:RC} displays the rotation curves in the left column for the small domain and the right column for the large domain. The top, middle, and bottom panels correspond to the initial data corresponding to our solutions with $M_{max} = 10$, 20, and 40, respectively. The black lines represent the rotation curve of {\it equilibrium} configurations, whereas the dashed lines are the average in time RCs of the {\it relaxed} configurations. At the bottom we show the relative deviation between the two. Notice also that the deviation is bigger for the bigger values of $M_{max}$. 

% --------------------------------
\subsection{Core-Halo scaling relation}

This relation between various scalars of a configuration is expressed in a relation between the core mass $M_{core}$, the total mass of the system, in our case that of the core-tail system $M=M_{max}$, and the total energy of the system $E$:

\begin{equation}
\frac{M_{core}}{M} = \beta ~ \Xi^{\alpha},
\label{eq:coremassrelation}
\end{equation}

\noindent where $\beta$ is some constant, $\Xi = \frac{|E|}{M^3}\left( \frac{\hbar}{Gm}\right)^2$ in physical units, and $\Xi=(4\pi)^2 \frac{|E|}{M^3}$ in code units. The exponent $\alpha$ has been found to have different values, depending on the method used to arrive at the core-tail configurations.

\begin{figure}
\includegraphics[width=8cm]{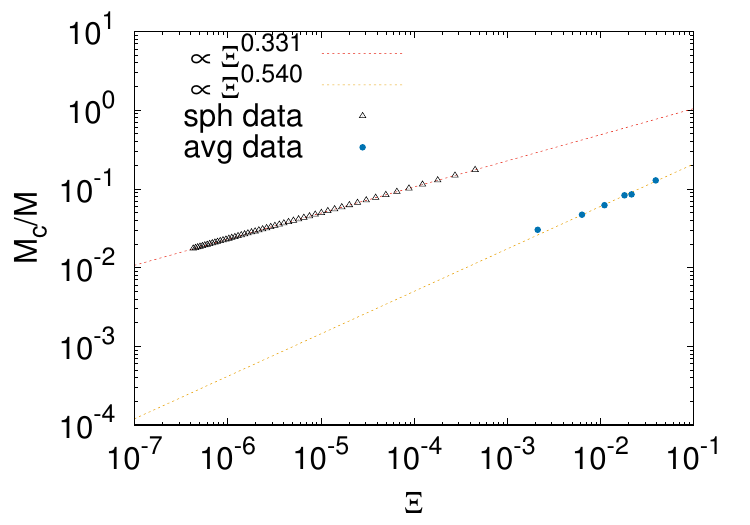}
\caption{Core-halo mass scaling relation for a number of equilibrium solutions, as well as for the relaxed solutions obtained from the six evolutions we have been working with. We fit the data with lines indicating that equilibrium configurations scale with $\alpha=1/3$, whereas relaxed solutions decay to a set that scales with $\alpha=0.54$.}
\label{fig:coremassrelation}
\end{figure}

The $CT$ solutions constructed in this paper have a scaling such that $\alpha\sim 1/3$ as illustrated in Figure \ref{fig:coremassrelation}. This behavior is consistent with the core-halo scaling relation found in \cite{Mocz:2017wlg} from the collision of many solitons randomly distributed and consistent also with the collision of many solitons under specific conditions in \cite{Luna2023}; it is also consistent with the exponent found in \cite{Nori2020,Mina_2022} resulting from the spherical collapse.

The core-halo scaling relation is expected to change for the relaxed solution as consequence of the initial kick of the velocity field of the tail on the core, as seen in Figure \ref{fig:diagnosis}. During the evolution the value of $M_{core}$ changes in time due to oscillations of the core and $M$ is domain size dependent because in the small domain $[-20,20]^3$ the portion of $M_{max}$ that fits within the domain is smaller than the portion in the big domain $[-40,40]^3$. Analogously, the integration of energy is also domain size dependent. The results in Figure \ref{fig:coremassrelation} for the relaxed configurations indicate the initial equilibrium configurations evolve towards a set of configurations that scale with $\alpha=0.54$ independently of the domain size. In order to compare with other analyses, we note that the range of $M_c/M$ values in Fig. \ref{fig:coremassrelation} is within $10^{-2}-10^{-1}$, one order of magnitud below the studies of multimergers in  \cite{Mocz:2017wlg} and \cite{Luna2023} based on multimerger scenarios, whereas our range intersects with values for binary merger based studies in \cite{Schwabe:2016}.
It is also important to note that this $\alpha$ holds for the scaling relations in e.g. \cite{Schwabe:2016,Mocz:2017wlg,Padilla_2021,Luna2023}, and is not the exponent of scaling relations in e.g. \cite{Schive:2014hza}, where $\alpha\sim 0.5$  from SFS, or in \cite{Chan_2022}, where also a different scaling relation is used.

% ---------------------------------------------
% ----->     SECTION.    <-----
% ---------------------------------------------
\section{Conclusions}
\label{sec:conclusions}

We have constructed core-tail solutions of the SP system with spherical symmetry. The family of solutions is parametrized by the total mass of the configuration $M_{max}$.

While the core is stationary, like ground state equilibrium solutions, the tail has a non-zero velocity. When these solutions evolve, they evolve and accommodate around a relaxed configuration that oscillates around a virialized state. The average density of these relaxed configurations preserves the core-tail structure although with a deviation from the profile of initial equilibrium solutions. This deviation is bigger for the bigger total mass of the system $M_{max}$. 

The motivation to construct these equilibrium solutions is that they may serve as realistic halo models, simpler than those constructed with large superposition of modes in \cite{YavetzLiHui2022} and much simpler than those obtained in SFS. However, between the initial equilibrium solutions and relaxed solutions, the latter seem to be the more consistent with those observed in SFS \cite{Schive:2014hza} and multi-soliton mergers \cite{Mocz:2017wlg,Veltmaat_2018}; unfortunately they need  evolution and thus are not as cheap anymore. This fact restricts the original optimistic applicability of easy-to-construct large catalogs of equilibrium solutions that could be used to fit rotation curves. Nevertheless, equilibrium solutions are similar to relaxed ones for small $M_{max}$, a regime where equilibrium solutions can be used due to their similarity to relaxed ones. In our examples, differences between equilibrium and relaxed configurations are small for $M_{max}$ as big as 10, a case where the core-halo mass ratio is of order $\sim 0.2$, namely, scenarios where the tail well dominates over the core. All cases with smaller $M_{max}$ include bigger core-halo mass ratios, meaning that the core dominates and therefore the influence of the tail is even smaller, which implies equilibrium and relaxed solutions are even more similar.

Finally, we also study the core-halo scaling relation. We found that equilibrium solutions scale with exponent $\alpha=1/3$, whereas the sample used to illustrate the evolution of equilibrium configurations decay into a set that scales with exponent $\alpha=0.54$.

% ----->     ACKNOWLEDGMENTS     <-----

\section*{Acknowledgments}
We express our gratitude to J. Luna Zagorac for providing valuable observations. This research is supported by grants CIC-UMSNH-4.9 and CONAHCyT Ciencias de Frontera Grant No. Sinergias/304001. The runs were carried out in the facilities of the IFM-UMSNH.

% -------------------------------------------------------
% -----     REFERENCES     ----------
% -------------------------------------------------------

\bibliography{BECDM}

\end{document}